\documentclass[twocolumn,showpacs,aps,floatfix,prd,nofootinbib]{revtex4}
\usepackage{graphicx}
\usepackage{dcolumn} 
\usepackage{amssymb} 
\usepackage{colordvi}
\usepackage{color}   
\usepackage{hhline}  
\usepackage{psfrag} 

\RequirePackage{xspace}

\newcommand{\BABARPubYear}    {07}
\newcommand{\BABARPubNumber}  {054}
\newcommand{\SLACPubNumber} {12806}
 
\usepackage{relsize}
\def\babar{\mbox{\slshape B\kern-0.1em{\smaller A}\kern-0.1em
    B\kern-0.1em{\smaller A\kern-0.2em R}}}
     
\mathchardef\Upsilon="7107
\def\Y#1S{\ensuremath{\Upsilon{(#1S)}}\xspace}

\def\pep2{PEP-II}

\mathchardef\Lambda="7103
\mathchardef\Sigma="7106
\def\Sigbar{\ensuremath{\kern 0.2em\overline{\kern -0.2em \Sigma}}{}}
\def\Lbar {\ensuremath{\kern 0.2em\overline{\kern -0.2em\Lambda\kern 0.05em}\kern-0.05em}{}}
\def\proton      {\ensuremath{p}\xspace}
\def\antiproton  {\ensuremath{\overline p}\xspace}
\def\B       {\ensuremath{B}\xspace}
\def\Bbar    {\kern 0.18em\overline{\kern -0.18em B}{}\xspace}
\def\qbar  {\ensuremath{\overline q}\xspace}
\def\q     {\ensuremath{q}\xspace}

\long\def\inst#1{\par\nobreak\kern 4pt\nobreak
  {\it #1}\par\vskip 10pt plus 3pt minus 3pt}
  
\begin{document}

\begin{flushleft}
SLAC-PUB-\SLACPubNumber \\
\babar-PUB-\BABARPubYear/\BABARPubNumber \\
\end{flushleft}

\title{\large \bf
\boldmath
Study of 
$e^+e^-\to \Lambda\Lbar,\;\Lambda\Sigbar^0,\;
\Sigma^0\Sigbar^0$ 
using Initial State Radiation with \babar\
}

\author{B.~Aubert}
\author{M.~Bona}
\author{D.~Boutigny}
\author{Y.~Karyotakis}
\author{J.~P.~Lees}
\author{V.~Poireau}
\author{X.~Prudent}
\author{V.~Tisserand}
\author{A.~Zghiche}
\affiliation{Laboratoire de Physique des Particules, IN2P3/CNRS et Universit\'e de Savoie, F-74941 Annecy-Le-Vieux, France }
\author{J.~Garra~Tico}
\author{E.~Grauges}
\affiliation{Universitat de Barcelona, Facultat de Fisica, Departament ECM, E-08028 Barcelona, Spain }
% ...
\collaboration{The \babar\ Collaboration}
\noaffiliation

\begin{abstract}
We study the 
$e^+e^-\to \Lambda\Lbar\gamma,\;
\Lambda\Sigbar^0\gamma,\;
\Sigma^0\Sigbar^0\gamma$ 
processes using 230 fb$^{-1}$ of integrated luminosity
collected by the \babar\ detector at $e^+e^-$ center-of-mass energy of 10.58 GeV.
From the analysis of the baryon-antibaryon mass spectra the cross 
sections for 
$e^+e^-\to \Lambda\Lbar,\; \Lambda\Sigbar^0,\;\Sigma^0\Sigbar^0$  
are measured in the dibaryon mass range from threshold up to 
3 GeV/$c^2$. The ratio of electric and magnetic form factors, 
$|G_E/G_M|$, is measured for $e^+e^-\to \Lambda\Lbar$, and limits
on the relative phase between $\Lambda$ form factors are obtained.
We also measure the $J/\psi\to \Lambda\Lbar,\;\Sigma^0\Sigbar^0$ and
$\psi(2S) \to \Lambda\Lbar$ branching fractions. 
\end{abstract}


\maketitle

\setcounter{footnote}{0} 

\section{Introduction\label{intro}}
\begin{figure}
\includegraphics[width=.3\textwidth]{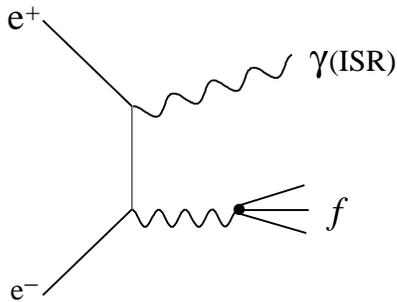}
\caption{The Feynman diagram describing the ISR process $e^+e^-\to f\gamma$, 
where $f$ is a hadronic system.
\label{diag3}}
\end{figure}

In this paper we continue the experimental study of baryon 
time-like electromagnetic form factors. In our previous work~\cite{BADpp} 
we have measured the energy dependence of the cross section for
$e^+e^-\to \proton\antiproton$ and of the proton form factor 
using the initial state radiation (ISR) technique.
Here we use this technique to study the processes 
\footnote{Throughout this paper the use of charge 
conjugate modes is implied.}
$e^+e^-\to \Lambda\Lbar,\; \Sigma^0\Sigbar^0,\; \Lambda\Sigbar^0$. 
The Born cross section for the ISR process
$e^+e^-\to f + \gamma$ (Fig.\ref{diag3}), where $f$ is a hadronic system,
integrated over the hadron momenta, is given by
\begin{equation}
\frac{{\rm d}\sigma_{e^+e^-\to f \gamma}(m)}
{{\rm d}m\,{\rm d}\cos{\theta_\gamma^\ast}} = 
\frac{2m}{s}\, W(x,\theta_\gamma^\ast)\,\sigma_{f}(m),
\label{eq1}
\end{equation}
where
$\sqrt{s}$ is the $e^+e^-$ center-of-mass energy (c.m.),
$m$ is the invariant mass of the hadronic system,
$\sigma_{f}(m)$ is the cross section for $e^+e^-\to f$ reaction,
$x\equiv{E_{\gamma}^\ast}/\sqrt{s}=1-{m^2}/{s}$, 
and $E_{\gamma}^\ast$ and $\theta_\gamma^\ast$
are the ISR photon energy and polar angle, respectively,  
in the $e^+e^-$ c.m. frame.
\footnote{Throughout this paper
the asterisk denotes quantities in the $e^+e^-$ c.m. 
frame.}
The function~\cite{BM}
\begin{equation}
W(x,\theta_{\gamma}^\ast)=
\frac{\alpha}{\pi x}\left(\frac{2-2x+x^2}{\sin^2\theta_{\gamma}^\ast}-
\frac{x^2}{2}\right)
\label{eq2}
\end{equation}
describes the probability of ISR photon emission for 
$\theta_{\gamma}^\ast\gg
m_e/\sqrt{s}$, where $\alpha$ is the fine structure constant and $m_e$ is the
electron mass. 
The cross section for the process $e^+e^- \to \B\Bbar$, 
where $B$ is a spin-1/2 baryon,
depends on magnetic ($G_M$) and electric ($G_E$)
form factors as follows:
\begin{equation}
\sigma_{\B\Bbar}(m) = \frac{4\pi\alpha^{2}\beta}{3m^2}
\left [|G_M(m)|^{2} + \frac{1}{2\tau}|G_E(m)|^{2}\right],
\label{eq4}
\end{equation}
where $\beta =\sqrt{1-4m_B^2/m^2}$ and $\tau=m^2/4m_B^2$; 
at threshold, $G_E=G_M$.
The cross section determines the
linear combination of the squared form factors
\begin{equation}
|F(m)|^2=\frac{2\tau|G_M(m)|^{2} + |G_E(m)|^{2}}
{2\tau+1},
\label{efform}
\end{equation}
and we define $|F(m)|$ to be the effective form factor~\cite{BADpp}.

The modulus of the ratio of electric and magnetic form factors
can be determined from the analysis of the distribution
of $\cos{\theta_B}$, where $\theta_B$ is the angle between the 
baryon momentum in the dibaryon rest frame and the momentum of 
the $\B\Bbar$ system in the $e^+e^-$ c.m. frame.
This distribution can be expressed as the sum of the terms proportional
$|G_M|^2$ and $|G_E|^2$. The $\theta_B$ dependencies of the
$G_E$ and $G_M$ terms are close to $\sin^{2}\theta_B$ and
$1+\cos^{2}\theta_B$ angular distributions for electric and         
magnetic form factors in the $e^+e^- \to \B\Bbar$ process.
The full differential cross section for
$e^+e^-\to \B\Bbar\gamma$~\cite{dkm} is given in the Appendix.

A nonzero relative phase between the electric and
magnetic form factors manifests itself in polarization
of the outgoing baryons. In the $e^+e^-\to \B\Bbar$ reaction
this polarization is perpendicular to the production plane~\cite{dub}. 
For the ISR process $e^+e^-\to \B\Bbar\gamma$
the polarization observables are analyzed in Refs.~\cite{kuhn_ll,dkm}.
The expression for the baryon polarization as a function of $G_E$, $G_M$,
and momenta of the initial electron, ISR photon, and final
baryon~\cite{dkm} is given in the Appendix. 
In the case of the $\Lambda\Lbar$
final state the $\Lambda\to \proton\pi^-$ decay can be used to
measure the $\Lambda$ polarization and hence the phase
between the form factors.

Experimental information on the $e^+e^-\to \Lambda\Lbar$,
$\Sigma^0\Sigbar^0$, $\Lambda\Sigbar^0$ reactions is very scarce.
The $e^+e^-\to \Lambda\Lbar$ cross section is measured as 
$100^{+65}_{-35}$ pb at 2.386 GeV, and at the same energy
upper limits for $e^+e^-\to \Sigma^0\Sigbar^0$
($ <120$ pb) and $e^+e^-\to \Lambda\Sigbar^0$ ($ < 75$ pb) 
cross sections have been obtained~\cite{DM2ll}.
No other experimental results exist.
\section{ \boldmath The \babar\ detector and data samples}                 
\label{detector}                                                           
We analyse a data sample corresponding to an integrated
luminosity of 230~fb$^{-1}$ recorded with                                                
the \babar\ detector~\cite{babar-nim} at the \pep2\                   
asymmetric-energy storage rings. At \pep2, 9-GeV electrons collide with     
3.1-GeV positrons at a center-of-mass energy of 10.58~GeV 
(the $\Upsilon$(4S) resonance). Additional data 
($\sim10\%$) recorded at 10.54 GeV are included in the
present analysis.

Charged-particle tracking is                                               
provided by a five-layer silicon vertex tracker (SVT) and                  
a 40-layer drift chamber (DCH), operating in a 1.5-T axial                 
magnetic field. The transverse momentum resolution                         
is 0.47\% at 1~GeV/$c$. Energies of photons and electrons                  
are measured with a CsI(Tl) electromagnetic calorimeter                    
(EMC) with a resolution of 3\% at 1~GeV. Charged-particle                  
identification is provided by specific ionization (${\rm d}E/{\rm d}x$)            
measurements in the SVT and DCH, and by an internally reflecting           
ring-imaging Cherenkov detector (DIRC). Muons are identified               
in the solenoid's instrumented flux return,                                
which consists of iron plates interleaved with resistive                   
plate chambers.                                          

Signal ISR processes are simulated with the Monte Carlo (MC) event generator
Phokhara~\cite{phokhara,kuhn_pp}. Because the polar-angle distribution of 
the ISR photon is peaked near $0^\circ$ and $180^\circ$, 
the MC events are generated with a restriction on the photon polar angle:
$20^\circ<\theta_{\gamma}^\ast<160^\circ$.
The Phokhara event generator includes next-to-leading-order radiative
corrections to the Born cross section. In particular, it
generates an extra soft photon emitted from the initial state. 
To restrict the maximum energy of the extra photon we require that 
the invariant mass of the dibaryon system and the ISR photon satisfies 
$M_{\B\Bbar\gamma}>8$ GeV/$c^2$. The generated events are subjected to detailed
detector simulation based on GEANT4~\cite{GEANT4}, 
and are reconstructed with the
software chain used for the experimental data. Variations in the detector
and in the beam background conditions are taken into account.
For the full simulation we use the differential cross section for 
the $e^+e^-\to \B\Bbar \gamma$ process with $G_E=G_M$.
In order to study angular distributions and model dependence of 
detection efficiency we produce two large samples of simulated
events at the generator level, one with $G_E=0$ and the other with
$G_M=0$, and reweight the events from the full simulation sample
according to the desired $|G_E/G_M|$ ratio.

Background from $e^+e^-\to \q\qbar$,
where $\q$ represents a $u$, $d$, $s$ or $c$ quark,
is simulated with the JETSET~\cite{JETSET} event generator. 
JETSET also generates  
ISR events with hadron invariant mass above 2 GeV/$c^2$ and 
therefore can be used to study ISR background with baryons in the 
final state. The most important background processes
$e^+e^-\to \B\Bbar\pi^0\gamma$, $e^+e^-\to \B\Bbar\pi^0$, and
$e^+e^-\to \Lambda\antiproton K^+\gamma$ are simulated separately. 
Three-body phase space and 
the Bonneau-Martin formula~\cite{BM} are used
to generate the angular and energy distributions for the final
hadrons and ISR photon, respectively. For these processes
extra soft-photon radiation from the initial state 
is generated using the structure function method~\cite{strfun}.

\section{\boldmath The reaction $e^+e^- \to \Lambda\Lbar\gamma$}
\begin{figure*}
\includegraphics[width=.33\textwidth]{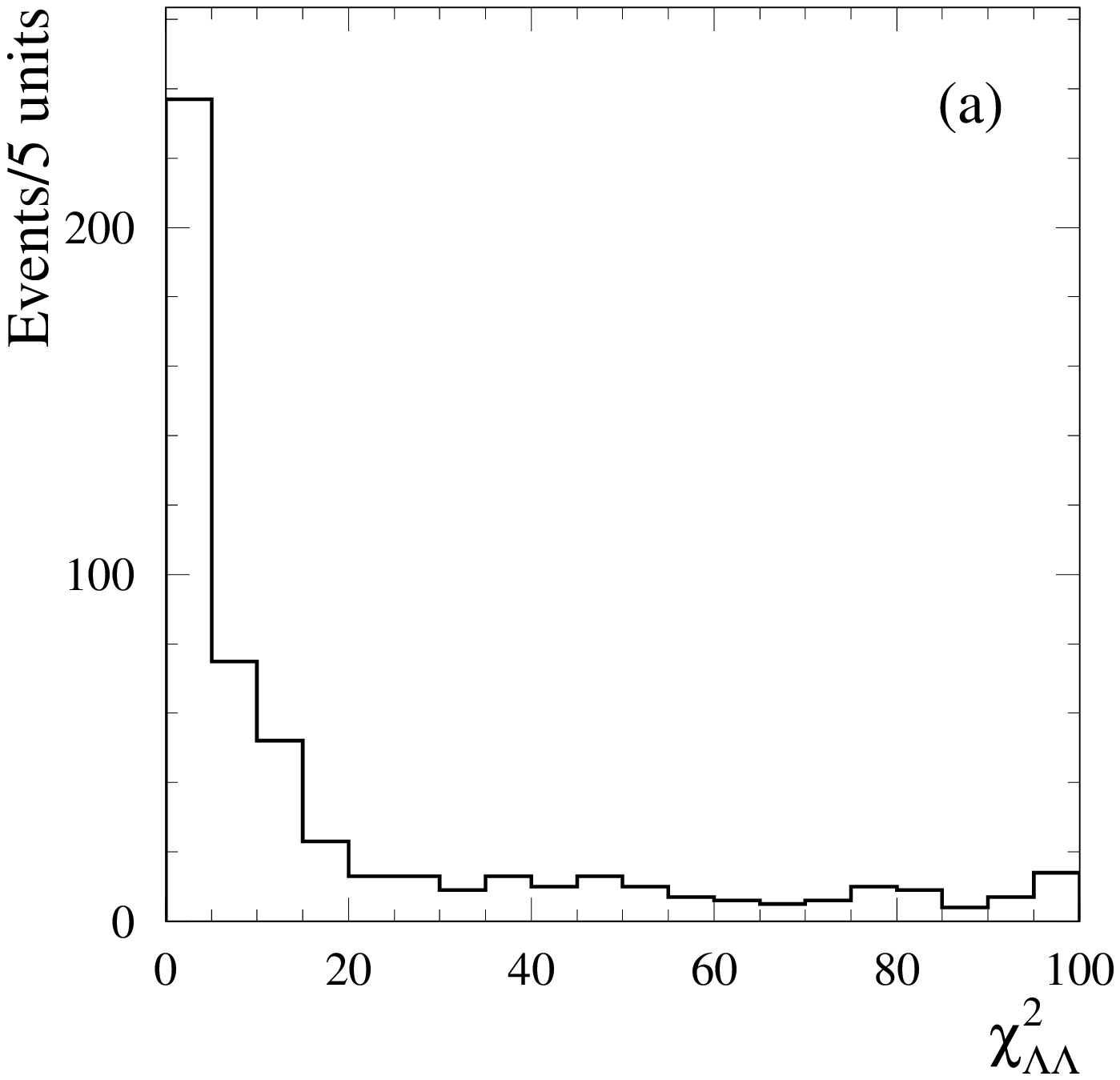}
\includegraphics[width=.33\textwidth]{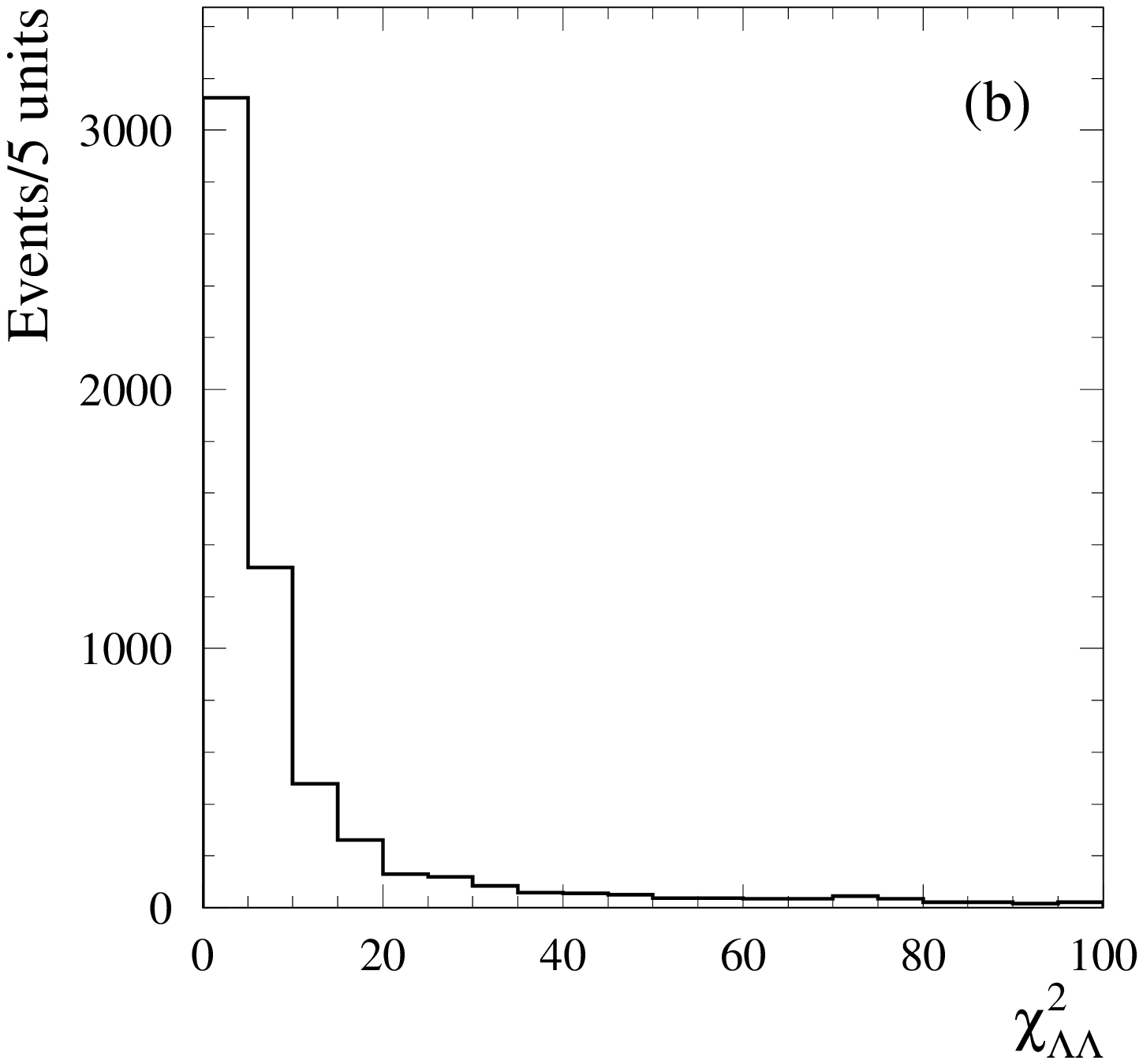}
\caption{The $\chi^2_{\Lambda\Lambda}$ distributions for 
data (a) and $e^+e^- \to \Lambda\Lbar\gamma$ simulation(b).
\label{dischi}}
\end{figure*}
\subsection{Event selection}
The initial selection of events requires the presence of
a high energy photon and at least one $\Lambda$ and one $\Lbar$ 
candidate. The hard photon must have energy in the c.m. frame
$E^\ast_\gamma>3$ GeV. The $\Lambda\to \proton\pi^-$
decay mode
with the branching fraction of $(63.9\pm0.5)\%$~\cite{pdg}
is used to identify $\Lambda$ 
candidates. Two oppositely-charged tracks 
are assigned the proton and pion mass hypotheses and fitted to a common vertex.
Any combination with invariant mass in the range 1.104-1.128 GeV/$c^2$ 
(the nominal $\Lambda$ mass is 1.115683(6) GeV/$c^2$~\cite{pdg}),
laboratory momentum greater than 0.5 GeV/$c$, and fit probability greater 
than 0.001 is considered a $\Lambda$ candidate. 
The candidate is then refitted with a $\Lambda$ mass constraint to 
improve the precision of the $\Lambda$ momentum measurement. 
To suppress combinatorial background we require that
at least one of the proton candidates be identified as 
a proton according to the specific ionization (${\rm d}E/{\rm d}x$)
measured in the SVT and DCH, and the Cherenkov angle measured              
in the DIRC.

For events passing the preliminary selection, 
we perform a kinematic fit that imposes energy and momentum conservation
at the production vertex to the $\Lambda$ and $\Lbar$
candidates and the photon with highest $E^\ast_\gamma$.
For events with more than one $\Lambda$ ($\Lbar$) candidate we 
consider all possible $\Lambda\Lbar$ combinations, and the one 
giving the lowest $\chi^2$ for the kinematic fit is retained.
The MC simulation does not accurately reproduce the shape
of the resolution function for the photon energy. This leads to a
difference in the  $\chi^2$ distributions resulting from the kinematic
fits to data and simulated events. To reduce this difference, only
the measured direction of the ISR photon is used in the fit;
its energy is a fit parameter.
\begin{figure*}
\begin{minipage}[t]{0.64\textwidth}
\includegraphics[width=.49\textwidth]{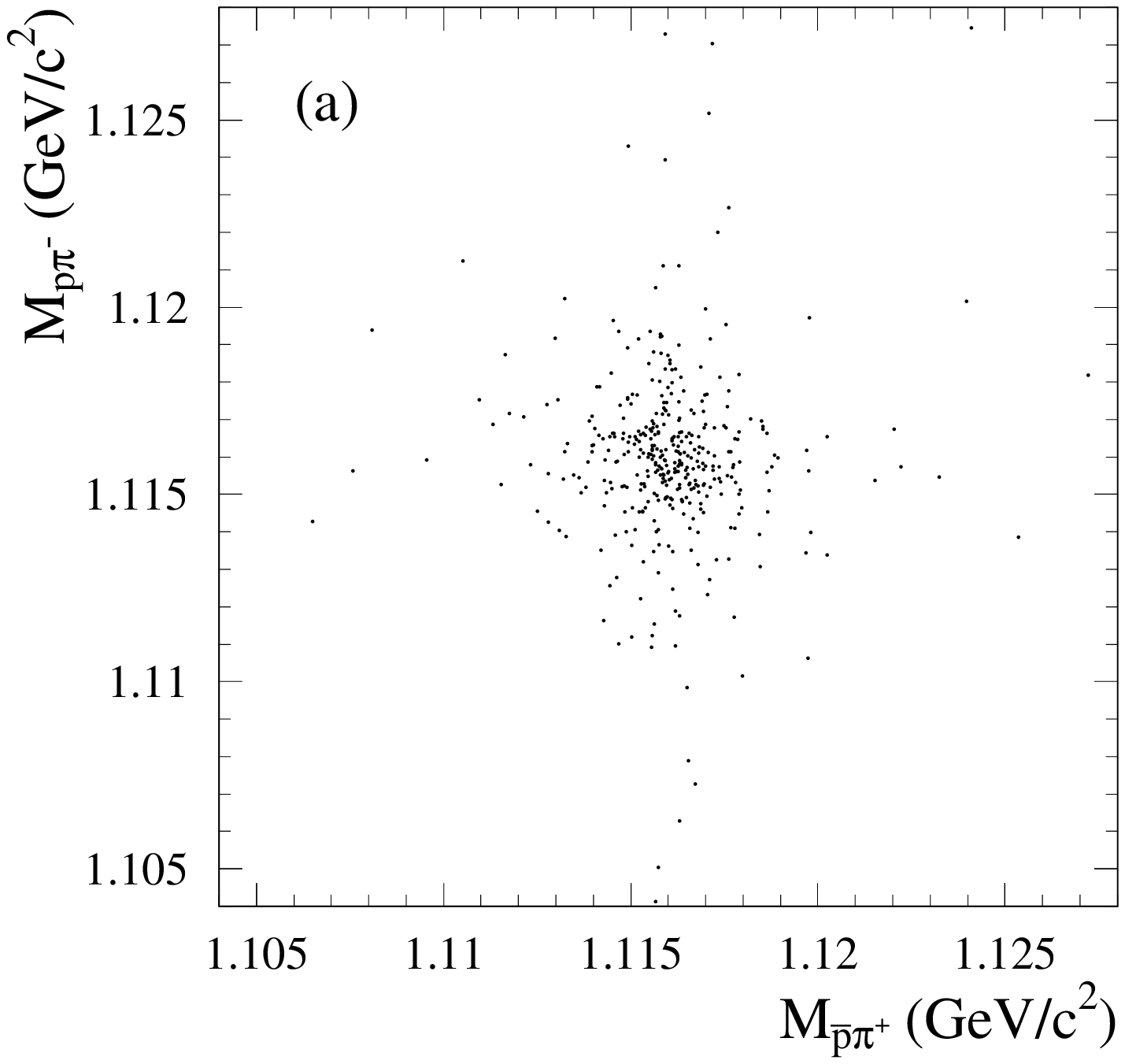}
\hfill
\includegraphics[width=.49\textwidth]{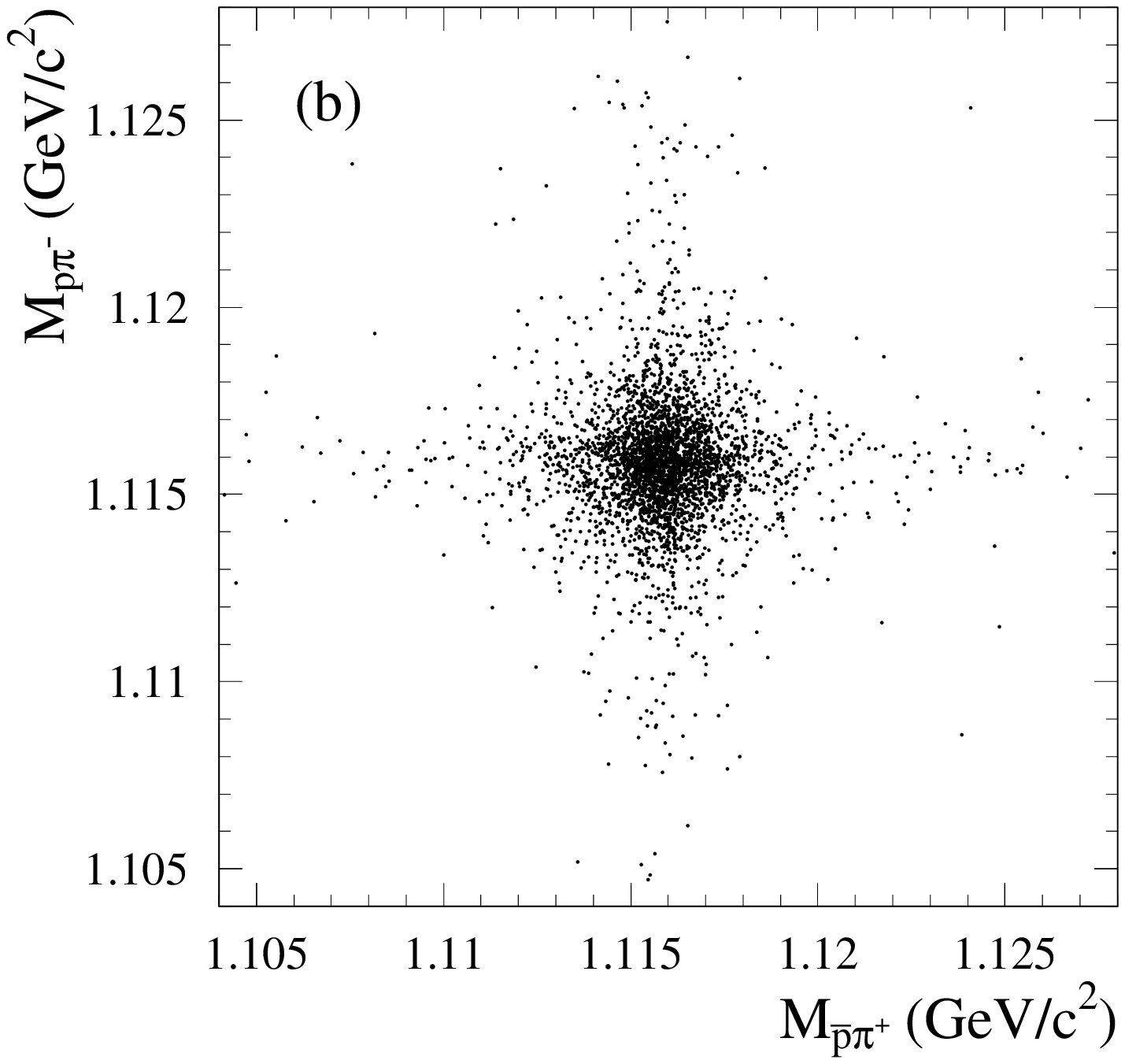}
\caption{Scatter plots of the invariant
mass of the $\Lambda$ candidate versus the invariant mass of the
$\Lbar$ candidate for data (a) and
$e^+e^- \to \Lambda\Lbar\gamma$ simulation (b).
\label{mlambda}}
\end{minipage}
\hfill
\begin{minipage}[t]{0.32\textwidth}
\includegraphics[width=.98\textwidth]{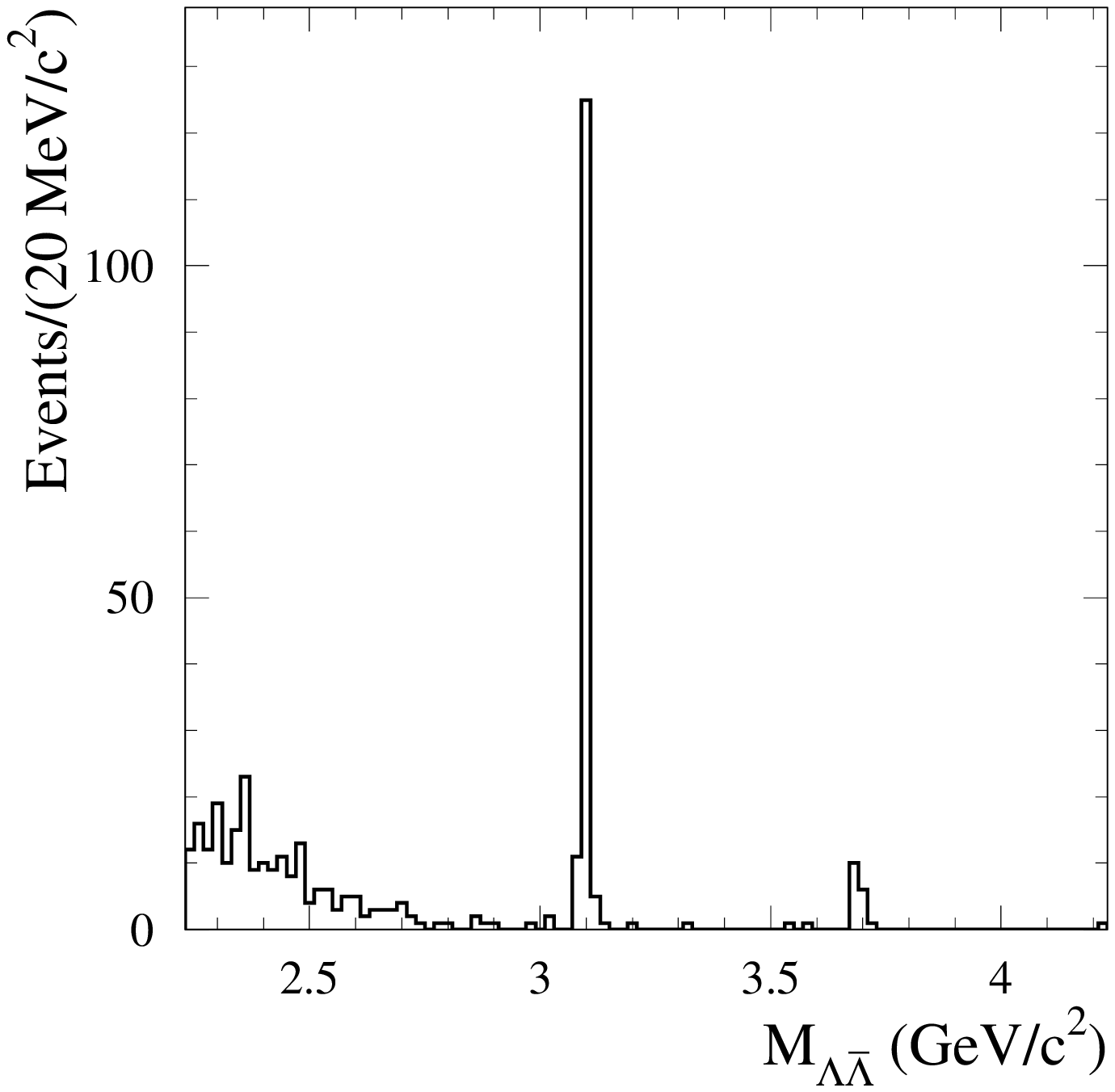}
\caption{The $\Lambda\Lbar$ mass spectrum for events satisfying
the $\Lambda\Lbar\gamma$ selection criteria.
\label{mlambda1}}
\end{minipage}
\end{figure*}
The $\chi^2$ distributions for the kinematic fit ($\chi^2_{\Lambda\Lambda}$)
to data events and to simulated $e^+e^- \to \Lambda\Lbar\gamma$
events are shown in Fig.~\ref{dischi}.
We select the events with $\chi^2_{\Lambda\Lambda}<20$ for further analysis.
The control region $20<\chi^2_{\Lambda\Lambda}<40$ is used for background
estimation and subtraction. 

Possible sources of background for the process under study are
those with only one $\Lambda$ in the final state, such as
$e^+e^-\to \Lambda\antiproton K^+\gamma$. Such events 
contain a charged kaon instead of one of the pion candidates.
To suppress this background we require that no charged pion candidate be
identified as a kaon.
This requirement rejects 70\% of the background from 
$e^+e^-\to \Lambda\antiproton K^+\gamma$ 
and only $\sim$2\% of signal events.

The scatter plot of the invariant
mass of the $\Lambda$ candidate versus the invariant mass of the
$\Lbar$ candidate for the 387 data events passing all the selection criteria
is shown in Fig.~\ref{mlambda}(a) and that for
simulated $e^+e^- \to \Lambda\Lbar\gamma$ events
is shown in Fig.~\ref{mlambda}(b). 
The $\Lambda\Lbar$ invariant mass spectrum for data events is shown in 
Fig.~\ref{mlambda1}. About half of the events have invariant mass below
3 GeV/$c^2$. Signals due to $J/\psi\to \Lambda\Lbar$ and 
$\psi(2S)\to \Lambda\Lbar$ decays are also clearly seen.

\subsection{Background subtraction}\label{background}
Processes of three kinds potentially contribute background 
to the $e^+e^-\to\Lambda\Lbar\gamma$ data sample, namely those
with zero, one and two $\Lambda$'s in the final state.

The composition of the one-$\Lambda$ background is studied using JETSET
simulation. This background is dominated by 
$e^+e^-\to \Lambda\antiproton K^+\gamma$ events. Other processes also           
contain a charged kaon in the final state.
The level of the one-$\Lambda$ background can be estimated from the 
fraction of data events rejected by the requirement
that no $\pi^+$ candidate be identified as a $K^+$.
For $M_{\Lambda\Lbar}<3$ GeV/$c^2$ this fraction is 3/224,
and we estimate that the one-$\Lambda$ background does not exceed 
1.6 events at 90\% confidence level (CL). 

\begin{figure}
\includegraphics[width=.33\textwidth]{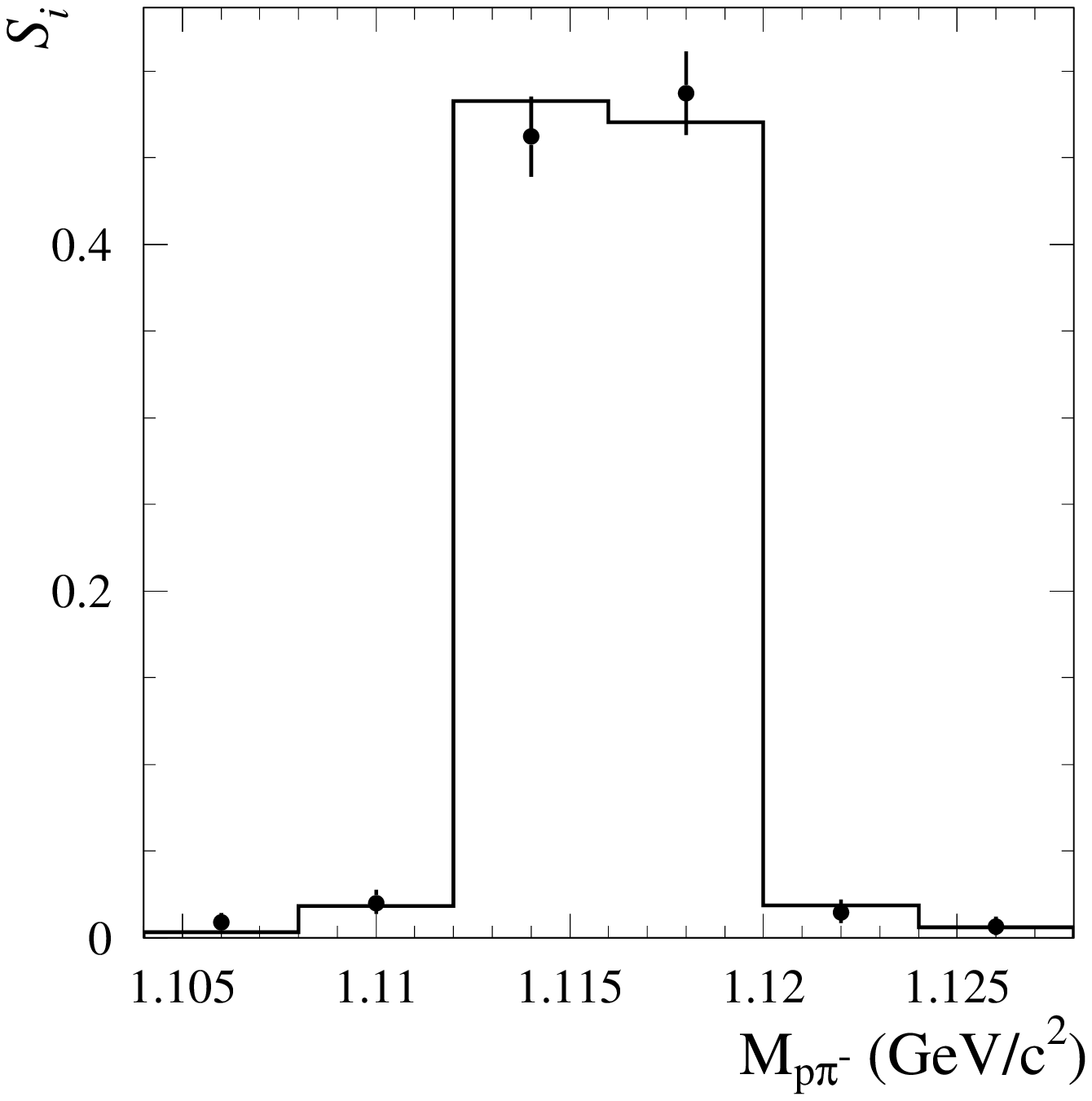}
\caption{The values of $S_i$ (see text) obtained from fits to data
(points with error bars) and from fits to $e^+e^- \to \Lambda\Lbar\gamma$ 
simulated events (histogram).
\label{si}}
\end{figure}
\begin{figure}
\includegraphics[width=.33\textwidth]{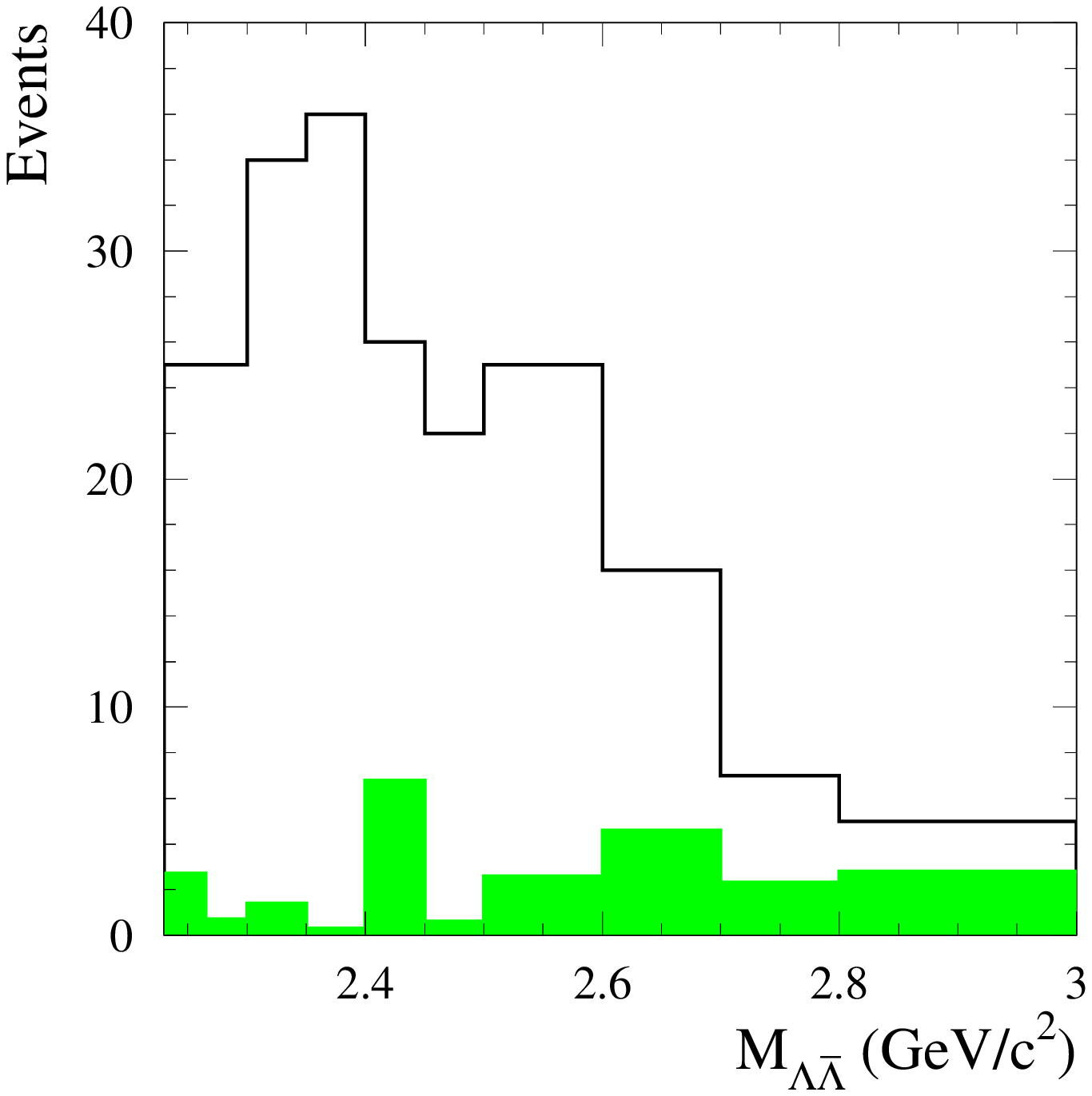}
\caption{The distribution of data events satisfying the                                                                
$\Lambda\Lbar\gamma$ selection criteria over chosen mass intervals.
The shaded histogram shows fitted background.
\label{mlambda3}}
\end{figure}
A more precise estimation 
(but based on JETSET prediction for the composition of one-$\Lambda$ events)
of this background is obtained
using a special selection
of $e^+e^-\to \Lambda\antiproton K^+\gamma$ events.
We select events with at least 4 charged tracks and a photon with 
$E_{\gamma}^\ast>3$ GeV. Two tracks, one of which is identified 
as a proton, must be combined to form a $\Lambda$ candidate, 
and the other two must originate from the $e^+e^-$ interaction
point and be identified as an antiproton and a positively-charged
kaon. For these events we perform the kinematic fit to the
$e^+e^-\to \Lambda\antiproton K^+\gamma$ hypothesis and require $\chi^2<20$.
The background for $e^+e^-\to \Lambda\antiproton K^+\gamma$ is estimated from 
the region $20<\chi^2<40$.
The total number of selected $\Lambda\antiproton K^+\gamma$ events is found to be 
$568\pm30$. Using the ratio of detection efficiencies for 
$\Lambda\Lbar\gamma$ and $\Lambda\antiproton K^+\gamma$ selections obtained 
from simulation, $(0.12\pm0.07)$\%, we calculate the number of 
$\Lambda\antiproton K^+\gamma$ events satisfying the $\Lambda\Lbar\gamma$ 
selection criteria to be $0.7\pm0.4$. Taking the ratio of 
$\Lambda\antiproton K^+\gamma$ to
all one-$\Lambda$ events (0.7) from JETSET simulation we
estimate the total number of one-$\Lambda$ background events 
to be  $1.0\pm0.6$.
The $e^+e^-\to \Lambda\antiproton K^+\gamma$ simulation is  reweighted to reproduce the
shape of the experimental $M_{\Lambda pK}$ distribution. The reweighted
events are then used to find the distribution of $M_{\Lambda\Lbar}$ for
events with only one real $\Lambda$.
We estimate $0.8\pm0.5$ background events to have $M_{\Lambda\Lbar} < 3$
GeV/$c^2$.

  The background processes with no real $\Lambda$'s are ISR processes
with four charged particles in the final state:
$e^+e^-\to 2\pi^+2\pi^-\gamma$,
$K^+K^-\pi^+\pi^-\gamma$,
$K^+K_S\pi^-\gamma$,
$\proton\antiproton\pi^+\pi^-\gamma$, $2\pi^+2\pi^-\pi^0$, etc.
The background from these processes can be estimated from an
analysis of the two-dimensional distribution of the $\Lambda$ and 
$\Lbar$ candidate mass values. The $6\times6$ two-dimensional
histogram corresponding to the plot in Fig.~\ref{mlambda}a 
is fitted by the following function:
\begin{equation}
n_{ij}=N_{2}S_iS_j+N_{0}B_{0i}B_{0j}+
N_{1}(S_iB_{1j}+S_jB_{1i})/2.
\label{eee}
\end{equation}
where the six mass intervals are those shown in Fig.~\ref{si}.
Here $N_{2}$, $N_{1}$, and $N_{0}$ represent the numbers of events with two, 
one, and zero $\Lambda$'s, respectively; $N_{2}$ and $N_{0}$ are free 
fit parameters, and $N_{1}$ is fixed at the value determined above
($0.8\pm0.5$ events for $M_{\Lambda\Lbar} < 3$ GeV/$c^2$); 
$S_i$ is the probability for a $\Lambda$
to have reconstructed mass in the $i$th mass bin, 
while $B_{0i}$ and $B_{1i}$ are the
probabilities for a false $\Lambda$ candidate from background 
with zero and one real $\Lambda$, respectively.
Since the one-$\Lambda$ background is small and
its presence leads to only small changes in the fitted $N_{2}$ and $N_{0}$ 
values, we use a uniform distribution for $B_{1}(m)$, i.e. all $B_{1i}=1/6$.
The $B_{0}(m)$ are parametrized by the linear function $B_{0i}=1/6+(i-3.5)\Delta$,
and five of the $S_i$ and $\Delta$ are free fit parameters. For 221 events with 
$M_{\Lambda\Lbar}<3$ GeV/$c^2$ the fit
yields $N_{2}=216\pm15$ and $N_{0}=4^{+4}_{-3}$. The fitted 
values of the $S_i$ are in good agreement with the values expected from 
$e^+e^- \to \Lambda\Lbar\gamma$ simulation (Fig.~\ref{si}).
In particular, 
$S_3+S_4=0.950\pm0.014$ for data and $0.953\pm0.003$ for simulation. 

The sources of two-$\Lambda$ background are processes with extra neutral
particle(s) in the final state:
$e^+e^- \to \Lambda\Lbar\pi^0$,
$e^+e^- \to \Lambda\Lbar\gamma\gamma$,
$e^+e^- \to \Lambda\Lbar\pi^0\gamma$, etc.
A significant fraction of $e^+e^- \to \Lambda\Lbar\pi^0$ events with
an undetected low-energy photon or with merged photons from $\pi^0$
decay are reconstructed under the $\Lambda\Lbar\gamma$ hypothesis 
with a low value of $\chi^2_{\Lambda\Lambda}$, and can not be separated
from the process under study. 
This background is studied by selecting a special subsample of events
containing a $\Lambda$ and a $\Lbar$ candidate and at least two
photons, one with energy greater than 0.1 GeV and the other with
c.m. energy above 3 GeV.
The two-photon invariant mass is required to be 
in the range 0.07 to 0.2 GeV/$c^2$. A kinematic fit to the 
$e^+e^- \to \Lambda\Lbar\gamma\gamma$ hypothesis is then performed.
Requirements on the $\chi^2$ ($\chi^2<20$) and
the two-photon invariant mass ($0.11<M_{\gamma\gamma}<0.16$  GeV/$c^2$)
are imposed to define $\Lambda\Lbar\pi^0$ candidates.
No data events satisfy these criteria, and the expected
background from $e^+e^- \to \Lambda\Lbar\gamma$ is $0.8\pm0.3$
events. The corresponding 90\% CL upper limit on the number of selected 
$\Lambda\Lbar\pi^0$ candidates is 1.6 events. 
Using the ratio of detection efficiencies
for $\Lambda\Lbar\pi^0$ and $\Lambda\Lbar\gamma$ selections 
($0.28\pm0.02$) we find that 
the $\Lambda\Lbar\pi^0$ background in $\Lambda\Lbar\gamma$ sample
does not exceed 6 events.
This upper limit is used as a measure of the                          
systematic uncertainty due to $\Lambda\Lbar\pi^0$ background.
We assume that the dibaryon mass spectrum in 
the $e^+e^- \to \Lambda\Lbar\pi^0$ process is similar
to that for the $\proton\antiproton\pi^0$ final state~\cite{BADpp}. In particular,
about 70\% of $\Lambda\Lbar\pi^0$ events are located
in the $\Lambda\Lbar$ mass region below 3 GeV/$c^2$. 
For this mass range this background
does not exceed 2\% of the selected $\Lambda\Lbar\gamma$ candidates.

\begin{table*}
\caption{The $\beta_i$ values obtained from simulation for 
signal and background processes, where
$\beta_i$ is the ratio of the number of selected $\Lambda\Lbar\gamma$
candidates with $20<\chi^2_{\Lambda\Lambda}<40$ to that with
$\chi^2_{\Lambda\Lambda}<20$. 
\label{beta}\\}
\begin{ruledtabular}
\begin{tabular}{lccccc}
&$\Lambda\Lbar\gamma$&$\Lambda\Sigbar^0\gamma$& 
$\Sigma^0\Sigbar^0\gamma$ & $\Lambda\Sigbar^0\pi^0$ & JETSET \\
$\beta_i$&$0.073\pm0.005$ & $0.83\pm0.07$ & $1.1\pm0.2$ &$0.81\pm0.09$ & $0.86\pm0.06$\\ 
\end{tabular}
\end{ruledtabular}
\end{table*}
\begin{table*}
\caption{$N$ is the number of selected $\Lambda\Lbar\gamma$ candidates
with $M_{\Lambda\Lbar}<3$ GeV/$c^2$, $N_{2s}$ is the number of signal events, 
$N_{0}$, $N_{1}$, $N_{2b}$ indicate the number of background events with zero, 
one, and two $\Lambda$'s in the final state, respectively, and 
$N_{\Lambda\Lbar\pi^0}$ is background from $e^+e^-\to\Lambda\Lbar\pi^0$.
\label{bkgsum}\\}
\begin{ruledtabular}
\begin{tabular}{lcccccc}
&$N$&$N_{2s}$&$N_{0}$&$N_{1}$&$N_{2b}$&$N_{\Lambda\Lbar\pi^0}$\\
$\chi^2_{\Lambda\Lambda}<20$&221&$204\pm19$&$4^{+4}_{-3}$&$0.8\pm0.5$&$12\pm10$&$<4$\\
$20<\chi^2_{\Lambda\Lambda}<40$&35&$15\pm3$&$9^{+7}_{-5}$&$0.6\pm0.4$&$11\pm8$&$<1$\\
$\chi^2_{\Lambda\Lambda}<20$ (JETSET)&522&$500\pm17$&$2.5\pm1.2$&$0.6\pm0.6$&$18\pm3$&$1.2\pm0.9$\\
\end{tabular}
\end{ruledtabular}
\end{table*}
The two-$\Lambda$ background other than from $e^+e^- \to \Lambda\Lbar\pi^0$
has the $\chi^2_{\Lambda\Lambda}$ distribution very different from
that for the process under study. Table~\ref{beta} shows 
the ratio of numbers of selected $\Lambda\Lbar\gamma$
candidates with $20<\chi^2_{\Lambda\Lambda}<40$ and 
$\chi^2_{\Lambda\Lambda}<20$
for signal and background processes. The ratios are obtained from simulation.
The column denoted ``JETSET'' shows the result of JETSET simulation for
background events containing two $\Lambda$'s in the final state.
From the number of selected two-$\Lambda$ events 
in the signal and control $\chi^2_{\Lambda\Lambda}$ regions, 
$N_2(\chi^2<20)$ and $N_2(20<\chi^2<40)$, 
the numbers of signal and background events with 
$\chi^2_{\Lambda\Lambda}<20$ can be calculated as:
\begin{eqnarray}
N_{2s}&=&\frac{N_2(\chi^2<20)-N_2(20<\chi^2<40)/\beta_{bkg}}
{1-\beta_{sig}/\beta_{bkg}},\nonumber\\
N_{2b}&=&N_2(\chi^2<20)-N_{2s},
\label{2bsub}
\end{eqnarray}
where $\beta_{bkg}$ is the ratio of fractions of events in the 
control and signal $\chi^2$ regions averaged over all processes
contributing into two-$\Lambda$ background. For this coefficient
we use $\beta_{bkg}=0.9\pm0.3$ which is close to the value obtained 
from the JETSET simulation, with an uncertainty covering the $\beta_{i}$ 
variations for different background processes. For the ratio for the 
signal process $\beta_{sig}$, we use the value obtained from simulation 
$\beta_{sig}=0.073\pm0.010$. The 
quoted error takes into account MC statistics, 
the data-MC simulation difference in $\chi^2$ distribution, and 
the $\beta_{sig}$ variation as a function of $\Lambda\Lbar$ mass.  
The difference between data and simulated $\chi^2$
distributions was studied in Ref.~\cite{BADpp} using 
the process $e^+e^-\to\mu^+\mu^-\gamma$. 
The resulting values of $N_{2s}$ and $N_{2b}$ for 
$\Lambda\Lbar$ masses below 3 GeV/$c^2$ are listed in 
Table~\ref{bkgsum}. The total background in the signal $\chi^2$ region 
from the 
processes with zero, one, and two $\Lambda$'s in final state is about 10\%.
The last row of the table shows the JETSET prediction for signal and background
events in the signal $\chi^2$ region. The simulation overestimates
the signal yield, but can be used for qualitative estimation of
background level.

The procedure for background estimation and subtraction described above is 
applied in each of the twelve $\Lambda\Lbar$ mass intervals 
indicated in Table~\ref{lamlamt}.
Due to restricted statistics we fit the two-dimensional histogram of
$M_{\proton\pi^-}$ vs $M_{\antiproton\pi^+}$ using $3\times3$ bins, and 
fix the $S_i$ (see Eq.(\ref{eee})) at the values obtained from MC simulation.  
The histograms for signal and control $\chi^2$ regions are fitted 
simultaneously. The free fit parameters are $N_{0}$ and $\Delta$
for the two $\chi^2$ regions, $N_{2s}$, and $N_{2b}$.
Fig.~\ref{mlambda3} shows the distribution of selected events over the 
chosen mass intervals. The shaded histogram shows the background 
contribution obtained from the fit.
The resulting numbers of signal events are listed in Table~\ref{lamlamt},
where the quoted errors include the statistical errors and errors due to
uncertainties in the $\beta_{sig}$, $\beta_{bkg}$ and $S_i$ coefficients.
These coefficients are varied within their uncertainties during fitting.
For the mass ranges $3.2 < M_{\Lambda\Lbar} < 3.6$ GeV/$c^2$ and 
$3.8 < M_{\Lambda\Lbar} < 5.0$ GeV/$c^2$ where we do not see
evidence for a signal above background, 90\% CL upper limits 
on the number of signal events are listed. The mass regions near 
the $J/\psi$ and $\psi(2S)$ will be considered separately 
in Sec.~\ref{jpsipp}.

\subsection{\boldmath Angular distributions for $e^+e^-\to \Lambda\Lbar\gamma$}
The modulus of the ratio of the electric and magnetic
form factors can be extracted from an analysis of the distribution of
$\cos{\theta_\Lambda}$, where $\theta_\Lambda$ is
the angle between the $\Lambda$ momentum in
the $\Lambda\Lbar$ rest frame and the momentum of the 
$\Lambda\Lbar$ system
in the $e^+e^-$ c.m. frame. This distribution is given by
\begin{eqnarray}
\frac{{\rm d}N}{{\rm d}\cos{\theta_\Lambda}}&=& 
A\,[ \rule{0mm}{5mm} H_M(\cos{\theta_\Lambda},M_{\Lambda\Lbar}) \nonumber\\
&+&\left |\frac{G_E}{G_M}\right|^2H_E(\cos{\theta_\Lambda},M_{\Lambda\Lbar})].
\label{an_fit}
\end{eqnarray} 
The functions $H_M(\cos{\theta_\Lambda},M_{\Lambda\Lbar})$ and 
$H_E(\cos{\theta_\Lambda},M_{\Lambda\Lbar})$
do not have an analytic form, and so are calculated using MC simulation.
To do this two samples of $e^+e^-\to \Lambda\Lbar\gamma$ events 
were generated, one with $G_E=0$ and the other with $G_M=0$, using 
generator level simulation. The angular dependencies of the 
resulting functions do not 
differ significantly from the $(1+\cos^2 \theta_\Lambda)$ and 
$\sin^{2}\theta_\Lambda$ functions corresponding to the magnetic and 
electric form factors in the case of $e^+e^-\to \Lambda\Lbar$.

\begin{figure*}
\includegraphics[width=0.33\textwidth]{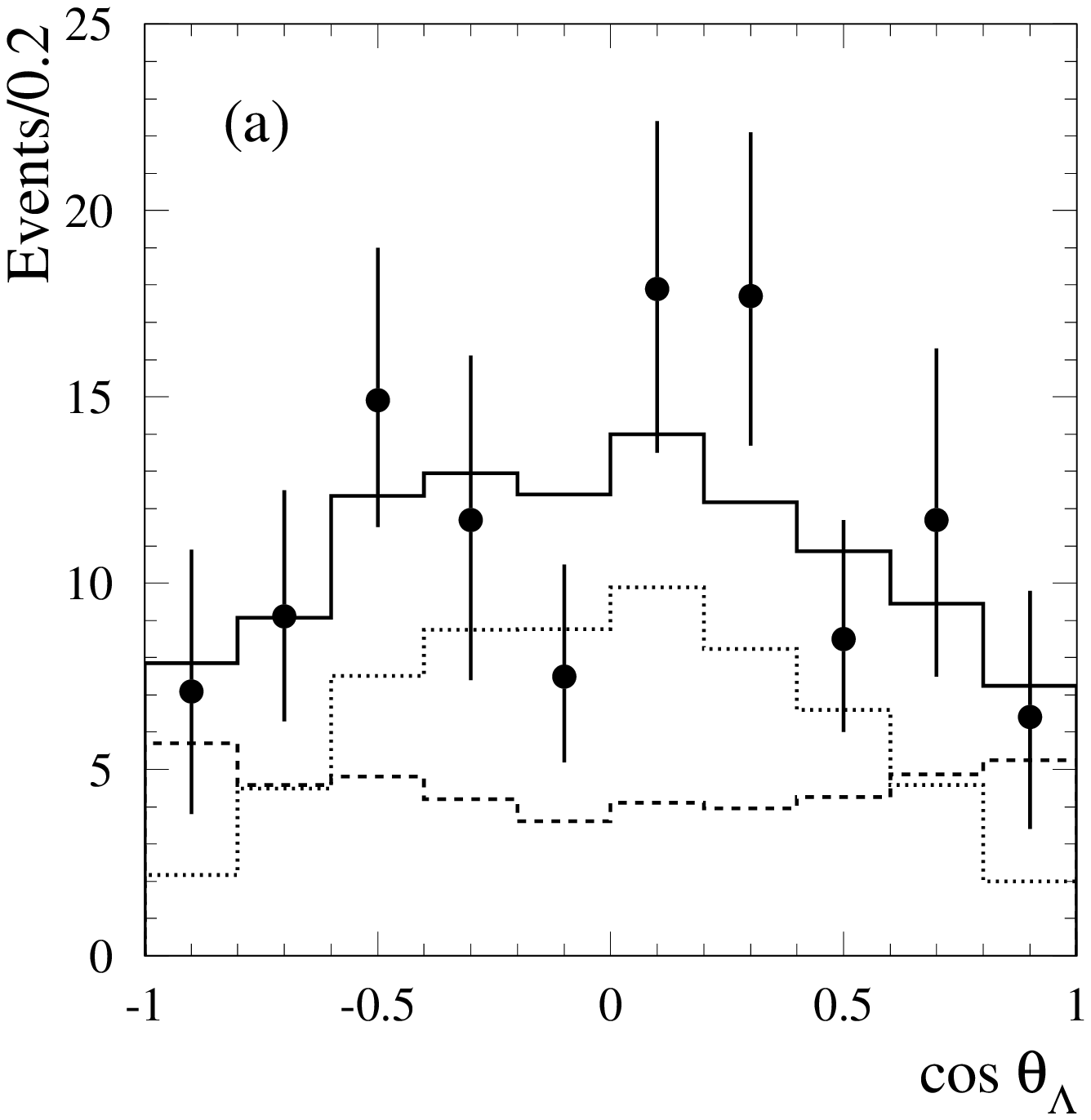}
\includegraphics[width=0.33\textwidth]{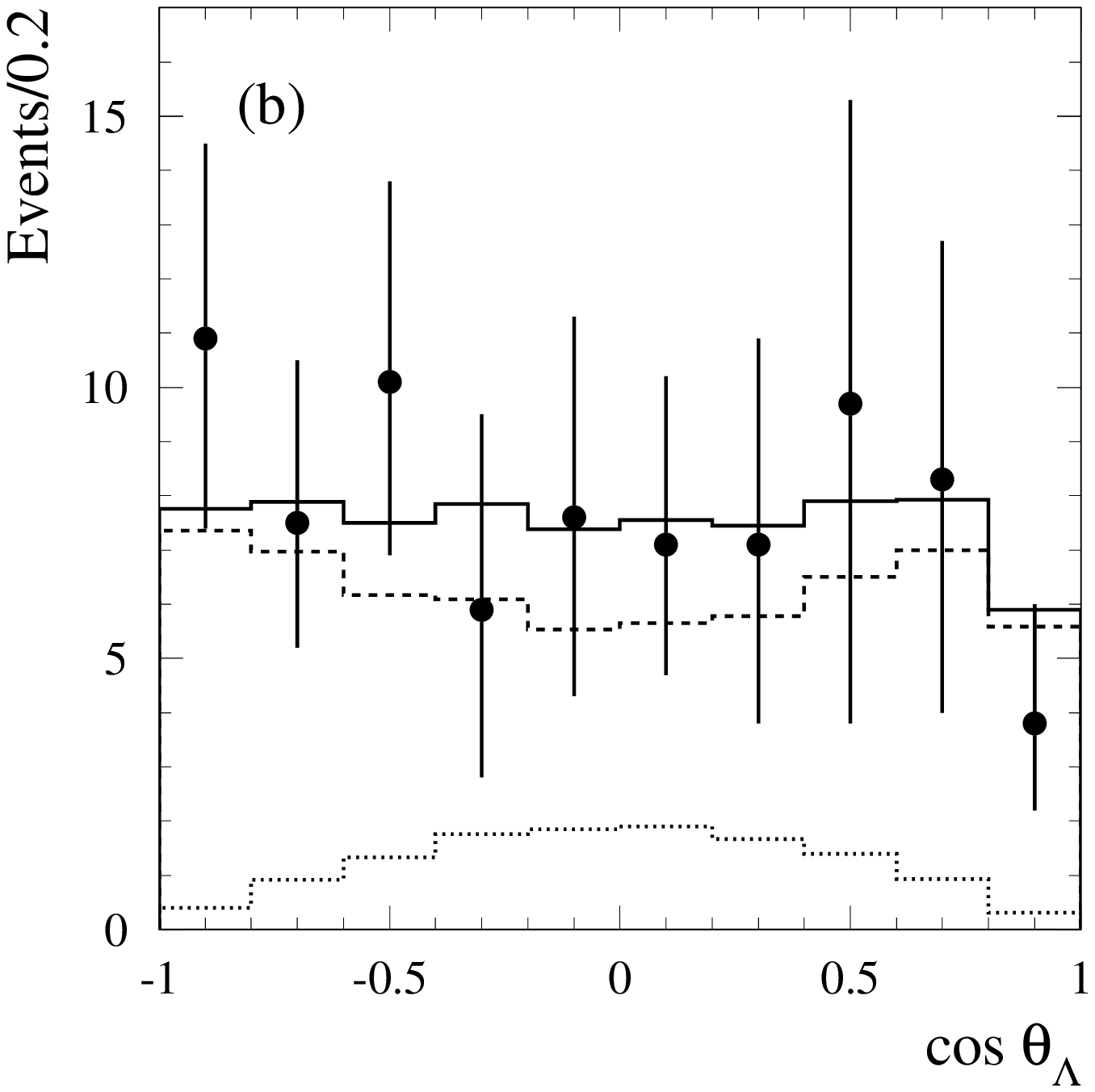}
\caption{The $\cos{\theta_\Lambda}$ distribution for
the mass regions 2.23--2.40 GeV/$c^2$ (a),
and 2.40--2.80 GeV/$c^2$ (b).
The points with error bars represent the data after
background subtraction.
The histograms are fit results: the dashed  histogram
shows the contributions corresponding to the magnetic
form factor; the dotted histogram shows the contribution
from the electric form factor, and the solid histogram is the
sum of these two.
\label{as}}
\end{figure*}

The observed angular distributions are fitted in two mass intervals:
from $\Lambda\Lbar$ threshold to 2.4 GeV/$c^2$ and
from 2.4 GeV/$c^2$ to 2.8 GeV/$c^2$.
For each mass and angular interval, the background is subtracted 
by means of the procedure described in the previous section. The
angular distributions obtained are shown in Fig.~\ref{as}.
\begin{figure}
\includegraphics[width=.33\textwidth]{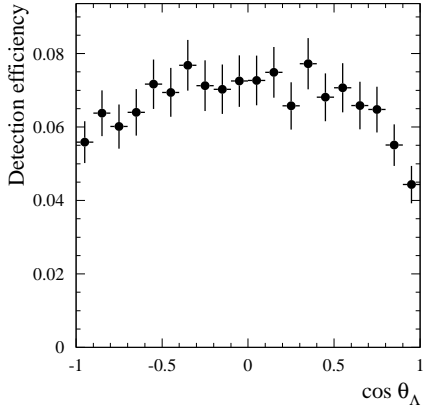}
\caption{The $\cos{\theta_\Lambda}$ dependence of the detection efficiency
for simulated events with $M_{\Lambda\Lbar}<2.8$ GeV/$c^2$.
\label{effvsang}}
\end{figure}
\begin{figure}
\includegraphics[width=.33\textwidth]{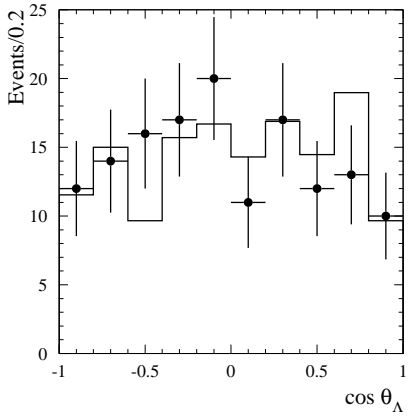}
\caption{The $\cos{\theta_\Lambda}$ distributions for
data (points with error bars) and simulation (histogram) corresponding 
to the reaction $e^+e^-\to J/\psi\gamma \to \Lambda\Lbar\gamma$.}
\label{aspsi}
\end{figure}
The distributions are fitted using the expression on the right-hand
side of Eq.~(\ref{an_fit}) with
two free parameters $A$ and $|G_E/G_M|$. The 
functions $H_M$ and $H_E$ are replaced by the histograms,
obtained from MC simulation with the $\Lambda\Lbar$ selection  
criteria applied. 
To take into account the effect of these criteria (Fig.~\ref{effvsang}),
the simulated events produced assuming $G_E=G_M$  
are reweighted according to the $\cos{\theta_\Lambda}$ distributions obtained
at generator level. These weight functions also
take account of the difference in $\Lambda\Lbar$ mass 
dependence between data and MC simulation. 
The histograms fitted to the angular 
distributions are shown in Fig.~\ref{as}. The following values of 
the $|G_E/G_M|$ ratio are obtained:
$$|G_E/G_M|=1.73^{+0.99}_{-0.57}\;\;\mbox{ (2.23--2.40 GeV)}/c^2,$$
$$|G_E/G_M|=0.71^{+0.66}_{-0.71}\;\;\mbox{ (2.40--2.80 GeV)}/c^2.$$
The quoted errors include both statistical and systematic uncertainties.
The net systematic uncertainty does not exceed 15\% of the statistical
error and includes the uncertainties due to background subtraction,
limited MC statistics, and the mass dependence of the $|G_E/G_M|$
ratio.

We also measure the angular distribution for $J/\psi\to\Lambda\Lbar$
decay, for which the shape is usually described by the form 
$(1+\alpha \cos^2{\theta})$. The world average value of 
$\alpha=0.65\pm0.10$~\cite{MARKll,DM2psi,BESll}. The distribution for 
$J/\psi\to \Lambda\Lbar$ decay in the present experiment is shown 
in Fig.~\ref{aspsi}. To remove background, this distribution was obtained
as the difference between
the histogram for the signal mass region (3.05-3.15 GeV/$c^2$) and that for 
the mass sidebands (3.00--3.05 and 3.15--3.20 GeV/$c^2$).
The data distribution is in  good agreement with that obtained from
simulation with $\alpha=0.65$.

Our results on the $|G_E/G_M|$ ratio are consistent both with 
$|G_E/G_M|=1$, valid at the $\Lambda\Lbar$ threshold, 
and with our results for the reaction $e^+e^-\to \proton\antiproton$
for which this ratio was found to be greater than unity near
threshold~\cite{BADpp}. The strong dependence of the $|G_E/G_M|$
ratio on the dibaryon mass near threshold is expected due to the
baryon-antibaryon final state interaction~\cite{FSI1,FSI2}.

\subsection{Mass dependence of the detection efficiency\label{efflam}}
\begin{figure}
\includegraphics[width=.33\textwidth]{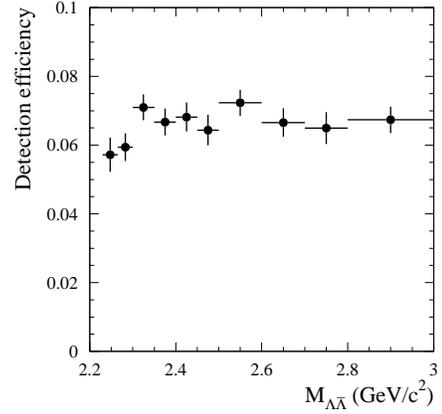}
\caption{The $\Lambda\Lbar$ mass dependence of detection
efficiency obtained from MC simulation.
\label{detef}}
\end{figure}
\begin{figure}
\includegraphics[width=.33\textwidth]{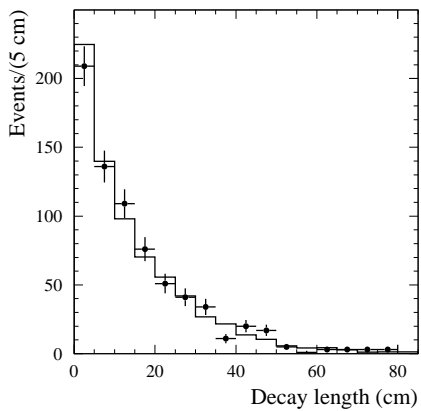}
\caption{The distribution of $\Lambda$ flight length for data
(points with error bars) and $e^+e^- \to \Lambda\Lbar\gamma$
simulation (histogram).
\label{fllen}}
\end{figure}
To first approximation the detection efficiency is determined
from MC simulation as the ratio of true $\Lambda\Lbar$ mass distributions
computed after and before applying the selection criteria. Since the 
$e^+e^-\to \Lambda\Lbar \gamma$ differential cross section depends
on two form factors the detection efficiency cannot be determined 
in a model-independent way.
We use a model in which the $|G_E/G_M|$ ratio is set to the values 
obtained from the fits to the experimental angular distributions 
for $M_{\Lambda\Lbar}<2.8$ GeV/$c^2$, and then set $|G_E/G_M|=1$ for 
higher masses. 
The detection efficiency obtained in this way is shown in
Fig.~\ref{detef}. This efficiency
includes the branching fraction for $\Lambda\to \proton\pi^-$ decay,
which is $(63.9\pm0.5)\%$~\cite{pdg}.
For $M_{\Lambda\Lbar}<2.8$ GeV/$c^2$ the variation
of the $|G_E/G_M|$ ratio within its experimental uncertainties 
leads to a 2.5\% model uncertainty. For higher masses, the model 
uncertainty is taken as half the difference between the detection
efficiencies corresponding $G_E=0$ and $G_M=0$; this yields a 5\%
uncertainty.

The efficiency determined from MC simulation 
($\varepsilon_{MC}$) must be corrected to account for
data-MC simulation differences in detector response:
\begin{equation}
\varepsilon=\varepsilon_{MC}\prod_i (1+\delta_i), 
\label{eq_eff_cor}
\end{equation}
where the $\delta_i$'s correct for the
several effects
discussed below, and summarized in Table~\ref{tab_ef_cor}.
\begin{table}
\caption{The values of the various efficiency corrections $\delta_i$
for the process $e^+e^-\to \Lambda\Lbar \gamma$. 
\label{tab_ef_cor}\\}                                
\begin{ruledtabular}
\begin{tabular}{ll}
effect                &$\delta_i$, (\%) \\
\hline
$\chi^2_{\Lambda\Lambda} < 20$       & $-2.0\pm2.0$  \\
no identified $K$                    & $-1.0\pm1.3$   \\
track reconstruction                 & $-1.0\pm3.8$  \\
$\antiproton$ nuclear interaction        & $+1.0\pm0.4$  \\
PID                                  & $+0.6\pm0.6$  \\
photon inefficiency                  & $-1.3\pm0.3$  \\
photon conversion                    & $+0.4\pm0.2$  \\
trigger                              & $-0.6\pm0.5$ for $M_{\Lambda\Lbar}<2.4 $ GeV/$c^2$ \\
\hline                                                                                                                    
total                 & $-3.9\pm4.6$ for $M_{\Lambda\Lbar}<2.4 $ GeV/$c^2$ \\
                      & $-3.3\pm4.6$ for $M_{\Lambda\Lbar}>2.4 $ GeV/$c^2$  \\
\end{tabular}                                                                                                            
\end{ruledtabular}
\end{table}        

The efficiency correction for the $\chi^2$ requirement was studied in 
Ref.~\cite{BADpp} for $e^+e^-\to \mu^+\mu^-\gamma$ and in 
Ref.~\cite{BAD4pi} for $e^+e^-\to 2\pi^+2\pi^-\gamma$. The corrections
were found to be $-(1.0\pm1.3)\%$ and $-(3\pm2)\%$, respectively.
For $e^+e^-\to \Lambda\Lbar\gamma$ 
we double the correction for $\mu^+\mu^-\gamma$, and assign 
a systematic uncertainty equal to the correction. 

The effect of requiring no identified $K$ is studied using
$e^+e^-\to J/\psi \gamma \to \Lambda\Lbar\gamma$ events.
The number of $J/\psi$ events is determined using the sideband subtraction 
method.
The event losses when requiring no identified $K$ are found to 
be $(2.1\pm1.2)\%$ in data and $(1.1\pm0.4)\%$ in MC simulation. 
The difference of these numbers is taken as the efficiency correction. 

Another source of data-MC simulation difference is track loss.
The correction due to the difference in track reconstruction is 
estimated to be -0.25\% per track with systematic uncertainty 
0.7\% for each proton and 1.2\% for each pion, which has
a softer momentum spectrum. Specifically, for the antiproton
track only, an extra systematic error originates from imperfect 
simulation of nuclear interactions of antiprotons in the 
detector material. This effect was studied in~\cite{BADpp}, and the 
corresponding efficiency correction is found to be $(1.0\pm0.4)\%$.
All corrections for track reconstruction described above were 
obtained for tracks originating from the $e^+e^-$ interaction point.
To estimate possible data-MC simulation difference due to $\Lambda$
flight path we compare the distributions of reconstructed
$\Lambda$ flight length (Fig.~\ref{fllen}). The data and
simulated distributions are in good agreement, and so there is no
need to introduce an extra efficiency correction for this effect.

The data-MC simulation difference for proton identification is
calculated using the $\proton/\antiproton$ identification probabilities for data and 
simulation obtained in Ref.~\cite{BADpp} for 
$e^+e^-\to J/\psi \gamma \to \proton\antiproton\gamma$. 

A correction must be also applied to the photon detection efficiency.
There are two main sources for this correction: data-MC simulation difference
in the probability of photon conversion in the detector material before the DCH,
and the effect of dead calorimeter channels. Both effects were studied
in Ref.~\cite{BADpp} using 
$e^+e^-\to \mu^+\mu^-\gamma$  and $e^+e^-\to \gamma\gamma$
events. 

The quality of the simulation of the trigger efficiency was also studied.
The overlap of the samples of events passing different trigger
criteria, and the independence of these triggers,
were used to measure the trigger efficiency.
A small difference ($-(0.6\pm0.5)\%$) in trigger efficiency between data and
MC simulation was observed for $\Lambda\Lbar$ masses below 2.4 GeV/$c^2$.

The total efficiency correction is $-(3.9\pm4.6)\%$ for 
$M_{\Lambda\Lbar}<2.4 $ GeV/$c^2$ and $-(3.3\pm4.6)\%$ for $M_{\Lambda\Lbar}>2.4 $ 
GeV/$c^2$.
The corrected detection efficiencies are listed in Table~\ref{lamlamt}.
The uncertainty in detection efficiency includes a simulation statistical
error, a model uncertainty, the error on the $\Lambda \to \proton\pi^-$ branching
fraction, and the uncertainty of the efficiency correction. 

\begin{table*}
\caption{The $\Lambda\Lbar$ invariant mass interval ($M_{\Lambda\Lbar}$),
net number of signal events ($N_s$), detection efficiency ($\varepsilon$), 
ISR luminosity 
($L$), measured cross section ($\sigma$), and effective form factor ($F$) 
for $e^+e^-\to \Lambda\Lbar$. The quoted errors on $\sigma$
are statistical and systematic, respectively. For the form factor, the 
total error is listed.
\label{lamlamt}\\}
\begin{ruledtabular}
\begin{tabular}{ccccccc}
$M_{\Lambda\Lbar}$ & $N_s$ & $\varepsilon$ &   $L$    & $\sigma$ & $|F|$ \\
(GeV/$c^2$)    &       &               & (pb$^{-1}$) &  (pb)    &       \\
\hline\\
2.23--2.27&$ 22.3^{+6.7}_{-6.5}$&$0.055\pm0.006$&  1.98&$ 204^{+62}_{-60}\pm22$  &$0.258^{+0.038}_{-0.044}$\\
2.27--2.30&$ 24.3^{+6.0}_{-5.8}$&$0.057\pm0.005$&  2.10&$ 202^{+50}_{-48}\pm18$  &$0.197^{+0.025}_{-0.027}$\\
2.30--2.35&$ 32.6^{+5.8}_{-5.2}$&$0.068\pm0.005$&  3.06&$ 155^{+28}_{-25}\pm12$  &$0.154^{+0.014}_{-0.014}$\\
2.35--2.40&$ 35.6^{+6.3}_{-6.3}$&$0.064\pm0.005$&  3.14&$ 176^{+31}_{-31}\pm15$  &$0.152^{+0.014}_{-0.016}$\\
2.40--2.45&$ 19.2^{+6.6}_{-6.4}$&$0.066\pm0.006$&  3.22&$  90^{+31}_{-30}\pm 8$  &$0.105^{+0.017}_{-0.020}$\\
2.45--2.50&$ 21.4^{+4.8}_{-4.3}$&$0.062\pm0.006$&  3.30&$ 104^{+24}_{-21}\pm10$  &$0.110^{+0.013}_{-0.013}$\\
2.50--2.60&$ 22.3^{+5.1}_{-4.5}$&$0.070\pm0.005$&  6.85&$  46^{+11}_{-9} \pm 4$  &$0.072^{+0.008}_{-0.008}$\\
2.60--2.70&$ 11.4^{+5.5}_{-5.9}$&$0.064\pm0.005$&  7.18&$  25^{+12}_{-13}\pm 2$  &$0.052^{+0.011}_{-0.016}$\\
2.70--2.80&$  4.7^{+4.0}_{-3.6}$&$0.063\pm0.006$&  7.52&$   10^{+9}_{-8} \pm 1$  &$0.033^{+0.012}_{-0.018}$\\
2.80--3.00&$  2.2^{+3.4}_{-3.7}$&$0.065\pm0.006$& 16.09&$2.1^{+3.2}_{-3.5}\pm0.2$&$0.016^{+0.009}_{-0.016}$\\
3.20--3.60&$ < 4.6$             &$0.055\pm0.005$& 39.88&$   < 2.1$               &$< 0.017$                \\
3.80--5.00&$ < 3.9$             &$0.066\pm0.005$&180.38&$   < 0.3$               &$< 0.009$                \\
\end{tabular}
\end{ruledtabular}
\end{table*}
\subsection{Cross section and form factor}\label{crosssec}
     The cross section for $e^+e^-\to \Lambda\Lbar$ is calculated from
the $\Lambda\Lbar$ mass spectrum using the expression
\begin{equation}         
\sigma(m)=\frac{({\rm d}N/{\rm d}m)_{corr}}{\varepsilon\, R\,         
{\rm d}L/{\rm d}m},         
\label{ISRcs}
\end{equation}         
where 
$({\rm d}N/{\rm d}m)_{corr}$ is the mass spectrum corrected for
resolution effects,         
${{\rm d}L}/{{\rm d}m}$ is the so-called ISR differential         
luminosity, $\varepsilon$ is the detection efficiency as a function of mass,
and $R$ is a radiative correction factor accounting for the Born mass         
spectrum distortion due to emission of extra photons by the initial         
electron and positron. The ISR luminosity is calculated         
using the total integrated luminosity $L$ and the probability density         
function for ISR photon emission~(Eq.~(\ref{eq2})):         
\begin{equation}         
\frac{{\rm d}L}{{\rm d}m}=\frac{\alpha}{\pi x}\left(         
(2-2x+x^2)\ln\frac{1+C}{1-C}-x^2 C\right)\frac{2m}{s}\,L.         
\label{ISRlum}         
\end{equation}
Here 
$C=\cos{\theta_0^\ast}$,
and $\theta_0^\ast$ determines the range of polar angles 
of the ISR photon in the $e^+e^-$ c.m. frame:
$\theta_0^\ast<\theta_\gamma^\ast<180^\circ-\theta_0^\ast$.
In our case  $\theta_0^\ast$  is equal to 20$^\circ$,         
since we determine efficiency using simulation with         
$20^\circ<\theta_\gamma^\ast<160^\circ$. The values of ISR luminosity
integrated over the corresponding mass interval are listed in Table~\ref{lamlamt}.
  
The radiative correction factor $R$ is determined using Monte Carlo         
simulation (at the generator level, with no detector simulation).         
The $\Lambda\Lbar$ mass spectrum is generated using only         
the pure Born amplitude for the $e^+ e^- \to \Lambda\Lbar\gamma $         
process, and then using a  model with next-to-leading-order         
radiative corrections included. The radiative correction factor, 
evaluated as the ratio of the second spectrum to the first, 
is found to be practically independent of mass, with
an average value equal to 1.0035 for masses below 3 GeV/$c^2$.
It should be noted that the value of $R$ depends on the criterion applied
to the invariant mass of the $\Lambda\Lbar\gamma$ system.
The value of $R$ obtained in our case corresponds to the requirement 
$M_{\Lambda\Lbar\gamma} > 8$ GeV/$c^2$ used in our simulation. 
The theoretical uncertainty in the radiative correction calculation
is estimated to be less than 1\%~\cite{phokhara}.
The calculated radiative correction factor does not take into         
account vacuum polarization, and the contribution of the latter is
included in the measured cross section.

\begin{figure*}
\begin{minipage}[t]{0.32\textwidth}
\includegraphics[width=.98\textwidth]{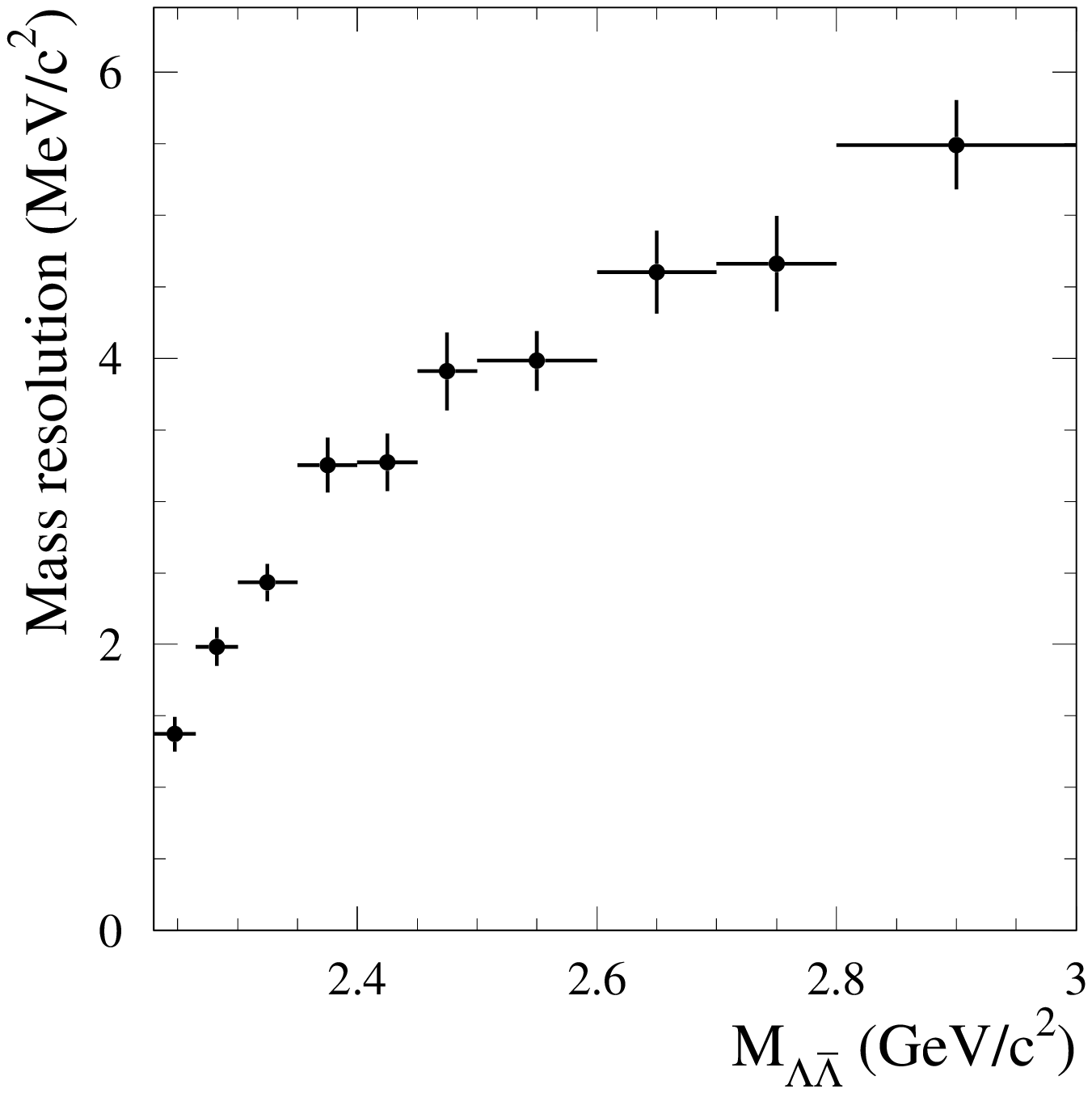}
\caption{The mass dependence of the mass resolution
calculated as the RMS of the $M_{\Lambda\Lbar}-M_{\Lambda\Lbar}^{true}$ 
distribution in MC simulation.
\label{res}}
\end{minipage}
\hspace{1mm}
\begin{minipage}[t]{0.32\textwidth}
\includegraphics[width=.98\textwidth]{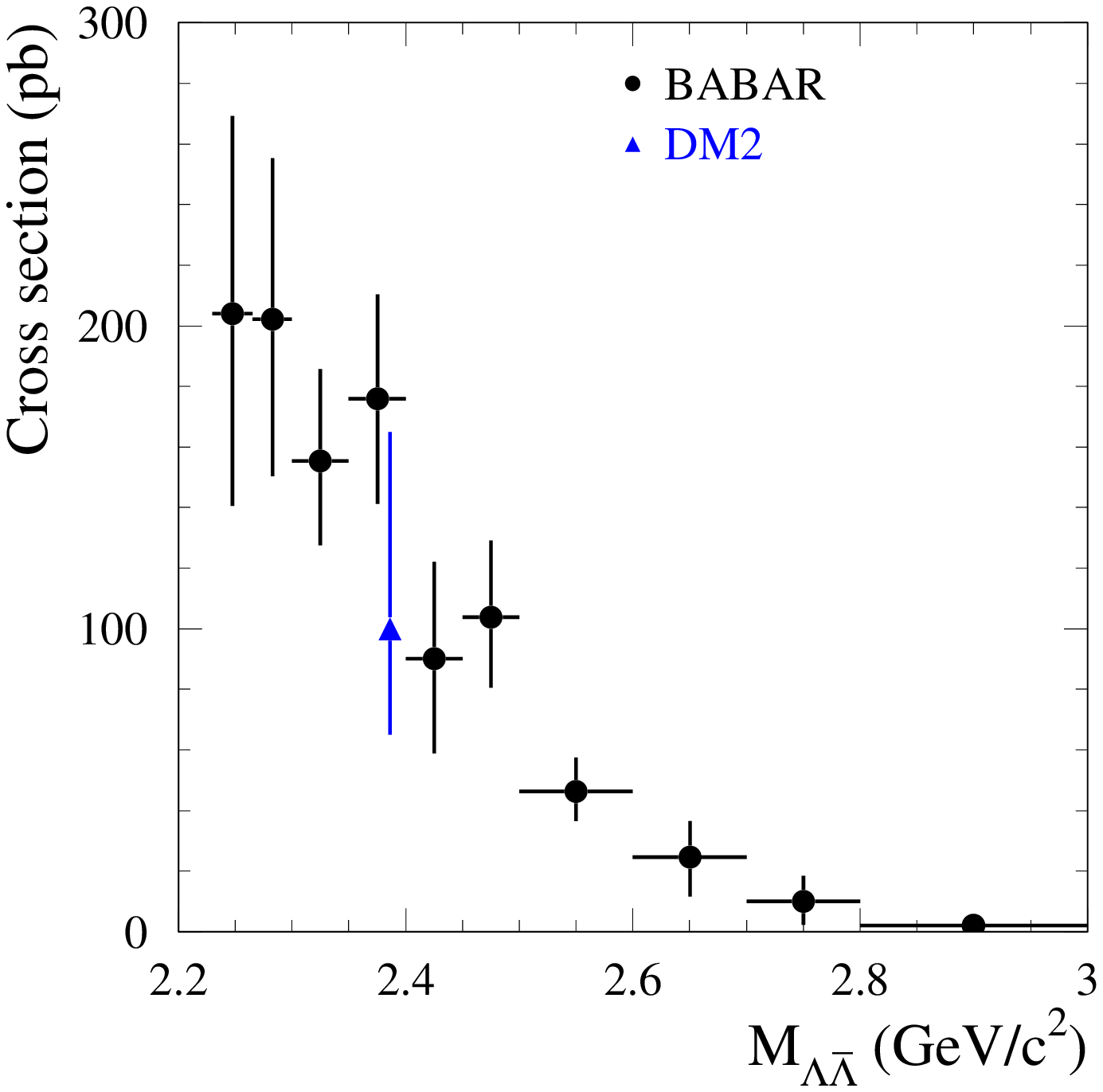}
\caption{The $e^+e^-\to \Lambda\Lbar$ cross section measured in 
the present experiment compared to the
DM2\cite{DM2ll} measurement.
\label{csb}}
\end{minipage}
\hspace{1mm}
\begin{minipage}[t]{0.32\textwidth}
\includegraphics[width=.98\textwidth]{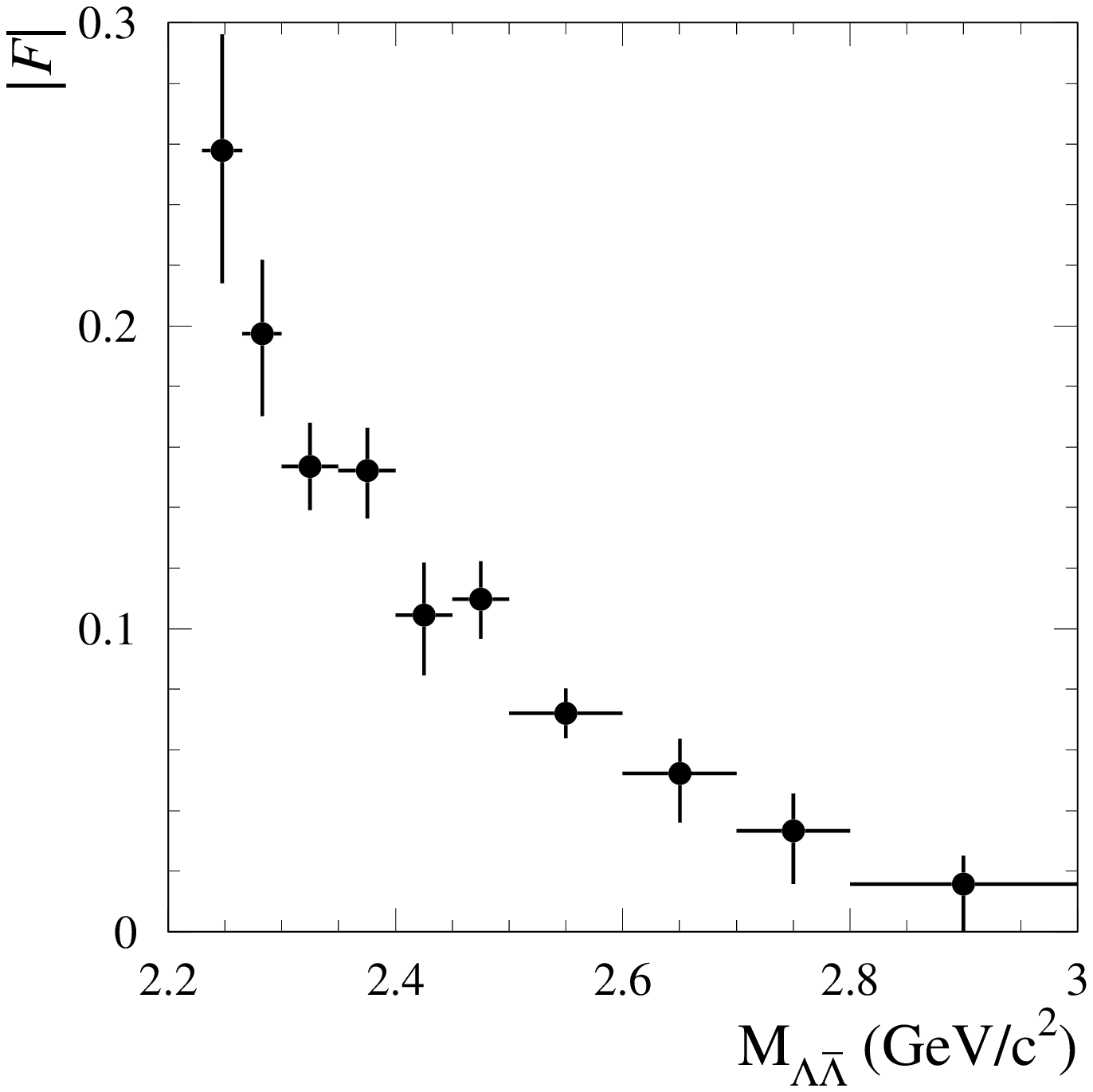}
\caption{The measured $\Lambda$ effective form factor.
\label{ff}}
\end{minipage}
\end{figure*}

The dependence of the mass resolution on the $\Lambda\Lbar$ invariant
mass is shown in Fig.~\ref{res}.
The mass resolution is calculated in simulation as the RMS deviation 
of the $M_{\Lambda\Lbar}-M_{\Lambda\Lbar}^{true}$ distribution. 
Since the chosen $M_{\Lambda\Lbar}$ intervals 
significantly exceed the                                     
mass resolution for all masses, we do not correct the mass spectrum for
resolution effects.

The measured cross section for $e^+e^-\to \Lambda\Lbar$ is shown in
Fig.~\ref{csb} and listed in Table~\ref{lamlamt}. 
The quoted errors are statistical and
systematic. The latter includes the systematic uncertainty in
detection efficiency, the uncertainty in total integrated luminosity
(1\%), and the uncertainty in the radiative correction (1\%).
The only previous measurement of the $e^+e^-\to \Lambda\Lbar$ cross
section, $100^{+65}_{-35}$ pb at 2.386 GeV~\cite{DM2ll}, is in agreement with
our results. 

The $e^+e^-\to \Lambda\Lbar$ cross section is a function of 
two form factors.
Due to the poorly determined $|G_E/G_M|$ ratio they cannot be extracted
from the data simultaneously with reasonable accuracy.
We introduce an effective form factor (Eq.(\ref{efform})) which is a
linear combination of $|G_E|^2$ and $|G_M|^2$.
The calculated effective form factor is shown in Fig.~\ref{ff}
and listed in Table~\ref{lamlamt}. 

\subsection{\boldmath $J/\psi$ and $\psi(2S)$ decays into $\Lambda\Lbar$}\label{jpsipp}
The  differential cross section for ISR production         
of a narrow resonance (vector meson $V$),         
such as $J/\psi$, decaying into the final state $f$ can be calculated         
using~\cite{ivanch}         
\begin{equation}         
\frac{{\rm d}\sigma(s,\theta_\gamma^\ast)}{{\rm d}\cos{\theta_\gamma^\ast}}
=\frac{12\pi^2 \Gamma(V\to e^+e^-) {\cal B}(V\to f)}{m_V\, s}\,         
W(s,x_0,\theta_\gamma^\ast),         
\label{eqpsi}         
\end{equation}         
where $m_V$ and $\Gamma(V\to e^+e^-)$ are the mass and electronic         
width of the vector meson $V$, $x_0 = 1-{m_V^2}/{s}$,         
and ${\cal B}(V\to f)$ is the branching fraction of $V$         
into the final state $f$. Therefore, the measurement of the number of
$J/\psi \to \Lambda\Lbar$ decays         
in $e^+ e^- \to \Lambda\Lbar\gamma$ determines the product of 
the electronic width and the branching fraction:         
$\Gamma(J/\psi \to e^+e^-){\cal B}(J/\psi \to \Lambda\Lbar)$.     
\begin{figure*}
\includegraphics[width=.33\textwidth]{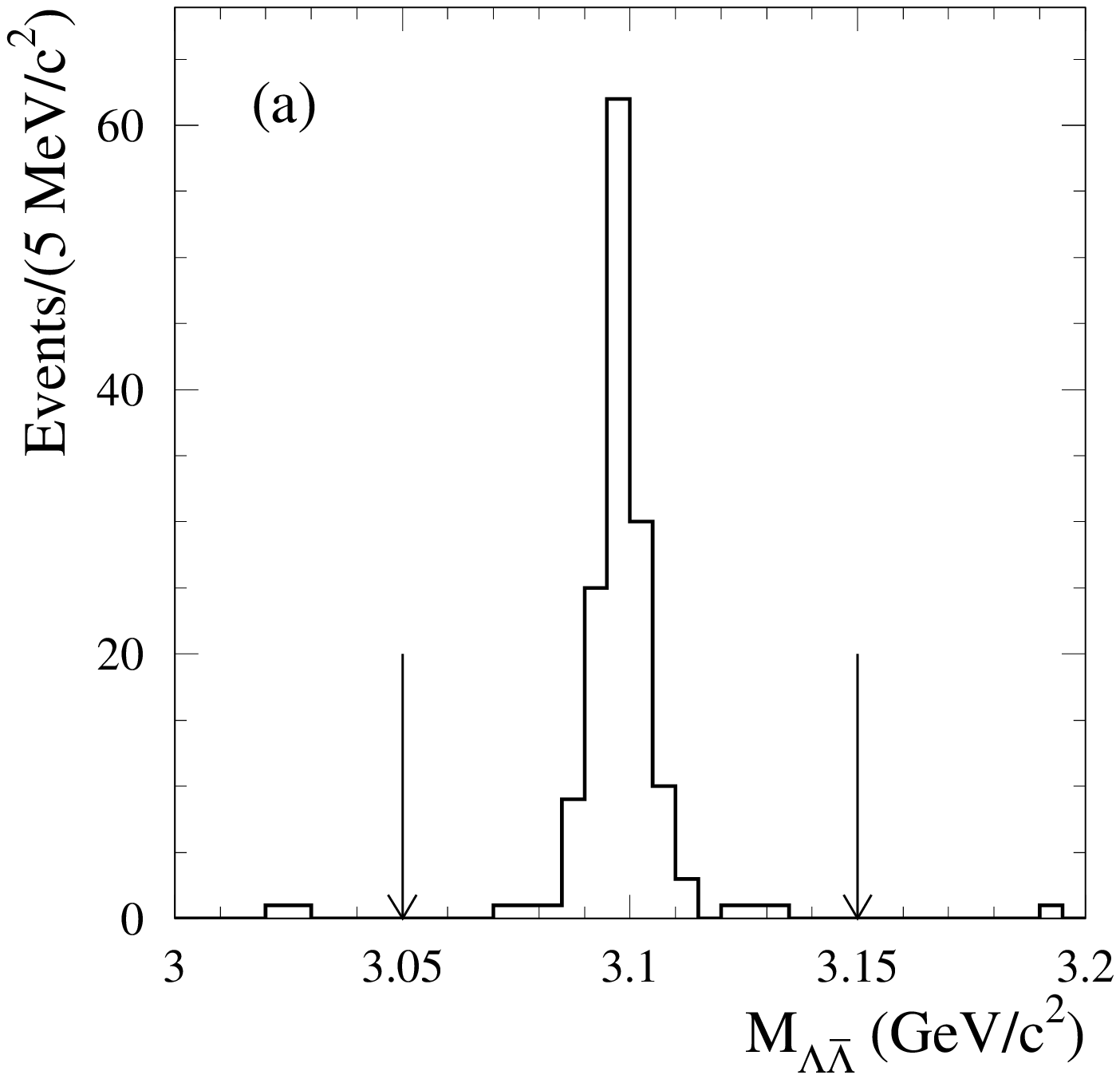}
\includegraphics[width=.33\textwidth]{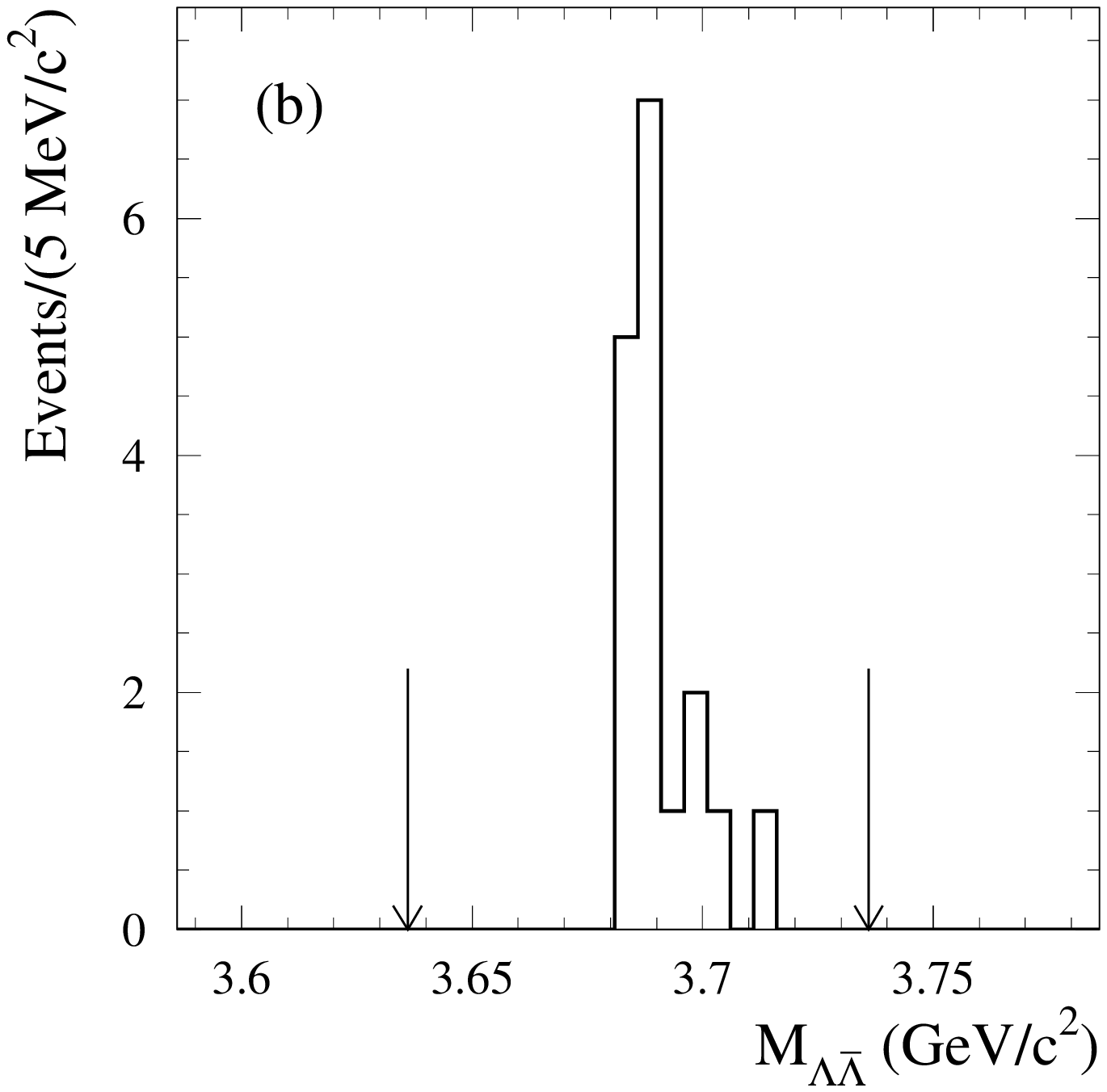}
\caption{The $\Lambda\Lbar$ mass spectra in the mass regions
near the $J/\psi$ (a) and the $\psi(2S)$ (b). The arrows
indicate the boundaries between the signal regions and sidebands.
\label{psiexp}}
\end{figure*}

The $\Lambda\Lbar$ mass spectra for selected events in the 
$J/\psi$ and $\psi(2S)$ mass regions are shown in Fig.~\ref{psiexp}.
We determine the number of resonance events by counting the events in
the signal region indicated in Fig.~\ref{psiexp}, and subtracting the
number in the two sidebands.
The following numbers of $J/\psi$ and $\psi(2S)$ events are obtained:
$N_{J/\psi}=142\pm 12$ and $N_{\psi(2S)}=17\pm4$.
A possible background due to $\psi\to \proton\antiproton\pi^+\pi^-$ decay
is estimated using the two-dimensional distribution of the masses of 
$\Lambda$ and $\Lbar$ candidates. It is found to be
$0.5^{+3.4}_{-0.5}$ events for $J/\psi$ and negligible for $\psi(2S)$.

The detection efficiency is estimated from MC simulation.
The event generator uses the experimental data for the angular
distribution of the $\Lambda$ in $J/\psi \to \Lambda\Lbar$ decay. 
This distribution is described by $1+\alpha \cos^2\theta$
with $\alpha=0.65\pm0.010$~\cite{MARKll,DM2psi,BESll}. For the $\psi(2S)$
the value $\alpha=0.69$ predicted in~\cite{Carimalo} is used. 
The error in the detection efficiency due to 
the uncertainty of $\alpha$ is negligible for the $J/\psi$ and
is taken to be 5\% for the $\psi(2S)$. 
The efficiencies corrected for data-MC simulation differences
are 0.062$\pm$0.004 for the $J/\psi$ and 0.059$\pm$0.005 for the 
$\psi(2S)$.

The cross section for         
$e^+e^-\to \psi\gamma\to \Lambda\Lbar\gamma$ for         
$20^\circ<\theta_\gamma^\ast<160^\circ$         
is calculated as
\begin{equation}
\sigma(20^\circ<\theta_\gamma^\ast<160^\circ)=\frac{N_{\psi}}         
{\varepsilon\, R\, L},
\label{eqpsi1}
\end{equation}
yielding $(9.8\pm0.9\pm0.6)$ fb and $(1.2\pm0.3\pm0.1)$ fb
for the $J/\psi$ and the $\psi(2S)$, respectively.
The radiative-correction factor $R=\sigma/\sigma_{Born}$ is         
$1.007\pm0.010$ for the $J/\psi$ and $1.011\pm0.010$ for the $\psi(2S)$,
obtained from MC simulation at the generator level.

The total integrated luminosity for the data sample  is         
$(230\pm 2)$ fb$^{-1}$.         
From the measured cross sections and Eq.~(\ref{eqpsi}),         
the following products are determined:         
\begin{eqnarray}
\Gamma(J/\psi\to e^+e^-){\cal B}(J/\psi\to \Lambda\Lbar) = 
(10.7\pm 0.9\pm 0.7)\mbox{ eV},\nonumber\\
\Gamma(\psi(2S)\to e^+e^-){\cal B}(\psi(2S)\to \Lambda\Lbar) = 
(1.5\pm0.4\pm0.1)\mbox{ eV}.\nonumber
\end{eqnarray}
The systematic errors include the uncertainties in detection efficiency,
integrated luminosity, and the radiative correction.         
 
Using the world-average values of the electronic widths~\cite{pdg},
the $\psi\to\Lambda\Lbar$ branching fractions are calculated to be         
\begin{eqnarray}
{\cal B}(J/\psi\to \Lambda\Lbar)=(1.92\pm0.21)\times 10^{-3},\nonumber\\
{\cal B}(\psi(2S)\to \Lambda\Lbar)=(6.0\pm1.5)\times 10^{-4}.\nonumber
\end{eqnarray}
Both results are higher than the current world-average values~\cite{pdg}:
$(1.54\pm0.19)\times 10^{-3}$ and $(2.5\pm0.7)\times 10^{-4}$, 
but in reasonable agreement with the more precise recent measurements:
$(2.03\pm0.15)\times 10^{-3}$ by BES~\cite{BESll} and
$(3.33\pm0.25)\times 10^{-4}$ by CLEO~\cite{cleo2s} and BES~\cite{bes2s}.
\begin{figure*}
\begin{minipage}[t]{0.32\textwidth}
\includegraphics[width=.98\textwidth]{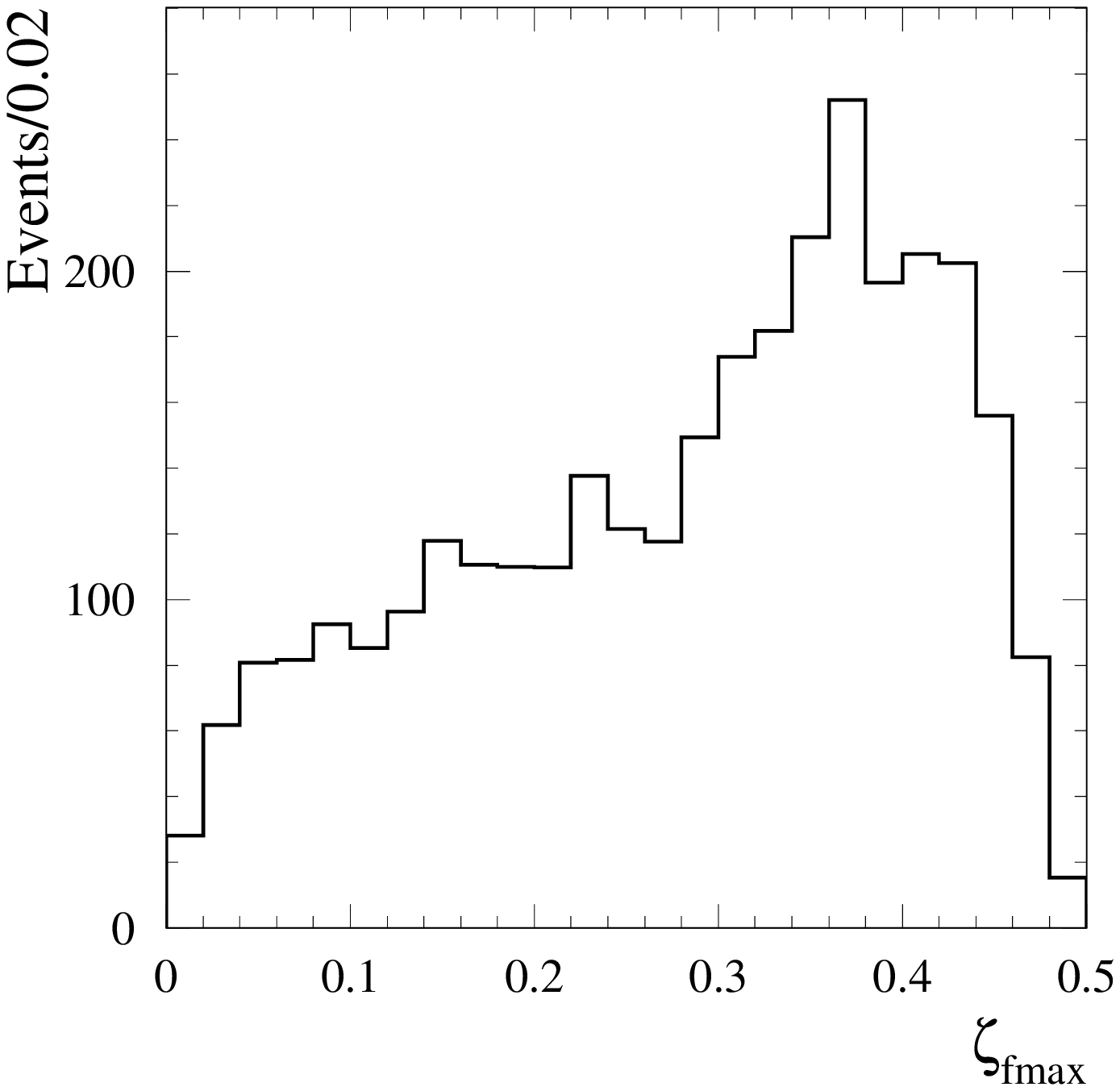}
\caption{The distribution of $\zeta_{fmax}$ for
selected simulated events for $e^+e^-\to \Lambda\Lbar\gamma$ with
$M_{\Lambda\Lbar} < 2.8$ GeV/$c^2$.\label{zeta}}
\end{minipage}
\hfill
\begin{minipage}[t]{0.64\textwidth}
\includegraphics[width=.49\textwidth]{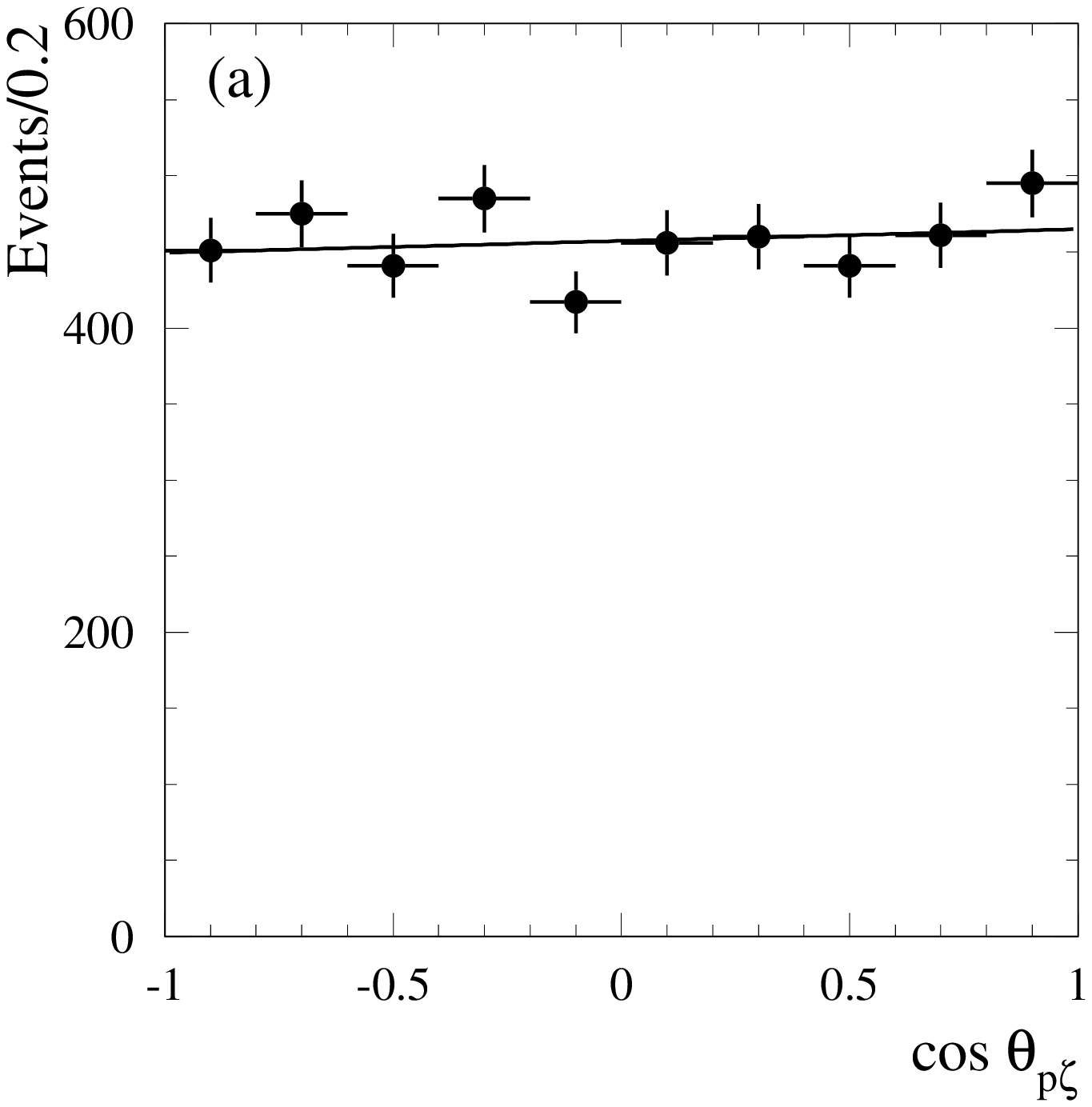}
\hfill
\includegraphics[width=.49\textwidth]{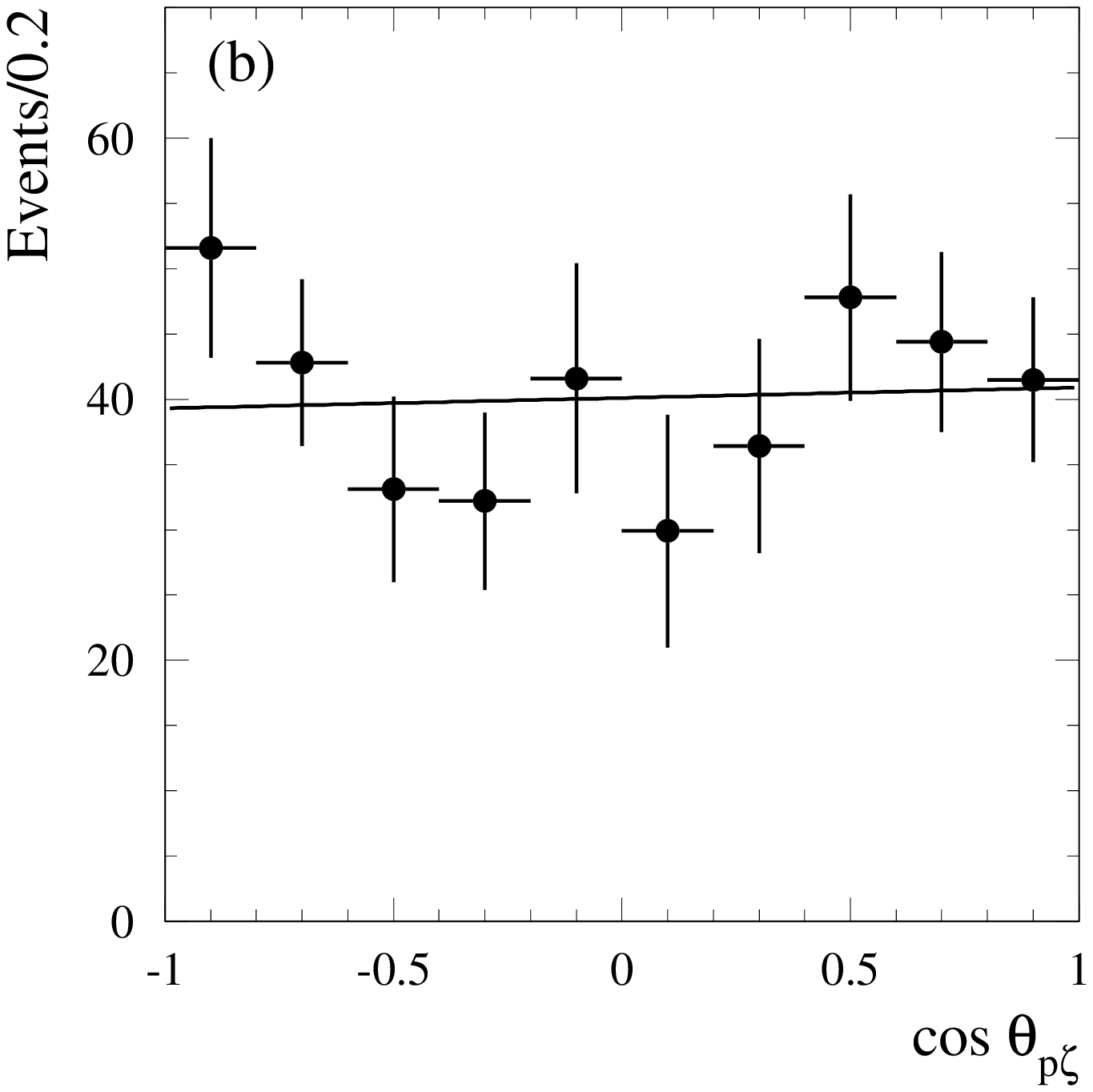}
\caption{The distribution of $\cos{\theta_{p\zeta}}$ for
selected $e^+e^-\to \Lambda\Lbar\gamma$ events with
$M_{\Lambda\Lbar} < 2.8$ GeV/$c^2$ in simulation (a) and 
in data (b).
\label{pol}}
\end{minipage}
\end{figure*}

\subsection{\boldmath Measurement of the $\Lambda$ polarization}
A nonzero relative phase $\phi$ between the electric and 
magnetic form factors leads to polarization
of the outgoing baryons. The exact formula for the $\Lambda$ ($\Lbar$)
polarization vector {\boldmath$\zeta_f$} is given in the Appendix.
The polarization is proportional to $\sin{\phi}$.
The magnitude of the polarization $\zeta_{fmax}=\zeta_f(\phi=\pi/2)$
calculated under the assumption that $|G_E|=|G_M|$
for simulated $e^+e^- \to \Lambda\Lbar\gamma$ events 
with $M_{\Lambda\Lbar} < 2.8$ GeV/$c^2$ is shown in Fig.~\ref{zeta}.
The simulated events were reweighted according to the $\Lambda\Lbar$ mass 
spectrum observed in data. The average value of $\zeta_{fmax}$ is equal 
to 0.285. The $\Lambda$  polarization can be measured using
the correlation between the direction of the $\Lambda$ polarization vector
and the direction of the proton from $\Lambda$ decay:
\begin{equation}
\frac{{\rm d}N}{{\rm d}\cos{\theta_{p\zeta}}} = 
A(1+\alpha_\Lambda \zeta_f \cos{\theta_{p\zeta}}), 
\end{equation}
where $\theta_{p\zeta}$ is the angle between the polarization axis
and the proton momentum in the $\Lambda$ rest frame, 
and $\alpha_\Lambda=0.642\pm0.013$~\cite{pdg}. For $\Lbar$,
$\alpha_{\Lbar}=-\alpha_\Lambda$.
The distribution of $\cos{\theta_{p\zeta}}$ for simulated 
$e^+e^- \to \Lambda\Lbar\gamma$ events
with $M_{\Lambda\Lbar} < 2.8$ GeV/$c^2$ 
(there is no $\Lambda$ polarization in the simulation)
is shown in Fig.~\ref{pol}(a).
We combine the $\Lambda$ and $\Lbar$ distributions taking
into account the different signs of $\alpha_\Lambda$ and 
$\alpha_{\Lbar}$. Since the distribution is flat, we conclude 
that there is no dependence of the detection efficiency on 
$\cos{\theta_{p\zeta}}$. 
A fit to the distribution using a linear function gives slope
consistent with zero.

The same distribution for data is shown in Fig.~\ref{pol}(b). 
In each angular interval the background is subtracted using the procedure
described in Sec.~\ref{background}. The distribution is fitted
using a linear function. The slope is found to be $0.020\pm0.097$.
The corresponding symmetric 90\% CL interval for $\Lambda$ polarization
averaged over the $\Lambda\Lbar$ mass range from threshold to
2.8 GeV/$c^2$ is
$$-0.22 < \zeta_f < 0.28.$$
Under the assumption $|G_E|=|G_M|$ ($\bar{\zeta}_{fmax}=0.285$), 
which does not contradict the data,
this interval can be converted to an interval for $\sin{\phi}$
as follows:
$$-0.76<\sin{\phi}<0.98.$$
Our statistics allow only very weak limits to be set on $\sin{\phi}$. 

\section{\boldmath The reaction $e^+e^- \to \Sigma^0\Sigbar^0\gamma$}
\subsection{Event selection\label{sigmasel}}
The $\Sigma^0$ hyperons are detected via the decay $\Sigma^0\to \Lambda\gamma$
(the branching fraction is 100\%~\cite{pdg}).
Therefore, the preliminary selection of  $e^+e^- \to \Sigma^0\Sigbar^0\gamma$
candidate events is similar to that for $e^+e^- \to \Lambda\Lbar\gamma$.
In addition, we require that an event contain at least two extra photons
with energy  greater than 30 MeV.
To suppress combinatorial background from events not containing 
two $\Lambda$'s in the final state, we apply a tighter selection criterion
on the mass of a $\Lambda$ ($\Lbar$) candidate: 
$1.110 < M_{\proton\pi^-} < 1.122$ GeV/$c^2$.

For events passing the preliminary selection, we perform a kinematic fit to
the $e^+e^- \to \Lambda\Lbar\gamma\gamma\gamma$
hypothesis. The photon with highest $E^\ast_\gamma$
is assumed to be the ISR photon. The fitted momenta of two other photons
and $\Lambda$-baryons are used to calculate $\Lambda\gamma$ and 
$\Lbar\gamma$ invariant masses.
For $\Sigma^0$ and $\Sigbar^0$ candidates these masses must to be in 
the range 1.155--1.23 GeV/$c^2$
(the nominal $\Sigma^0$ mass is 1.192642(24) GeV/$c^2$~\cite{pdg}).
We require that an event contain at least one $\Sigma^0$ and one $\Sigbar^0$
candidate. 
For events with more than three photons we iterate over all possible
photon combinations and find the one containing $\Sigma^0$ and 
$\Sigbar^0$ candidates and giving the lowest $\chi^2$ for the kinematic fit.

\begin{figure*}
\begin{minipage}[t]{0.32\textwidth}
\includegraphics[width=.98\textwidth]{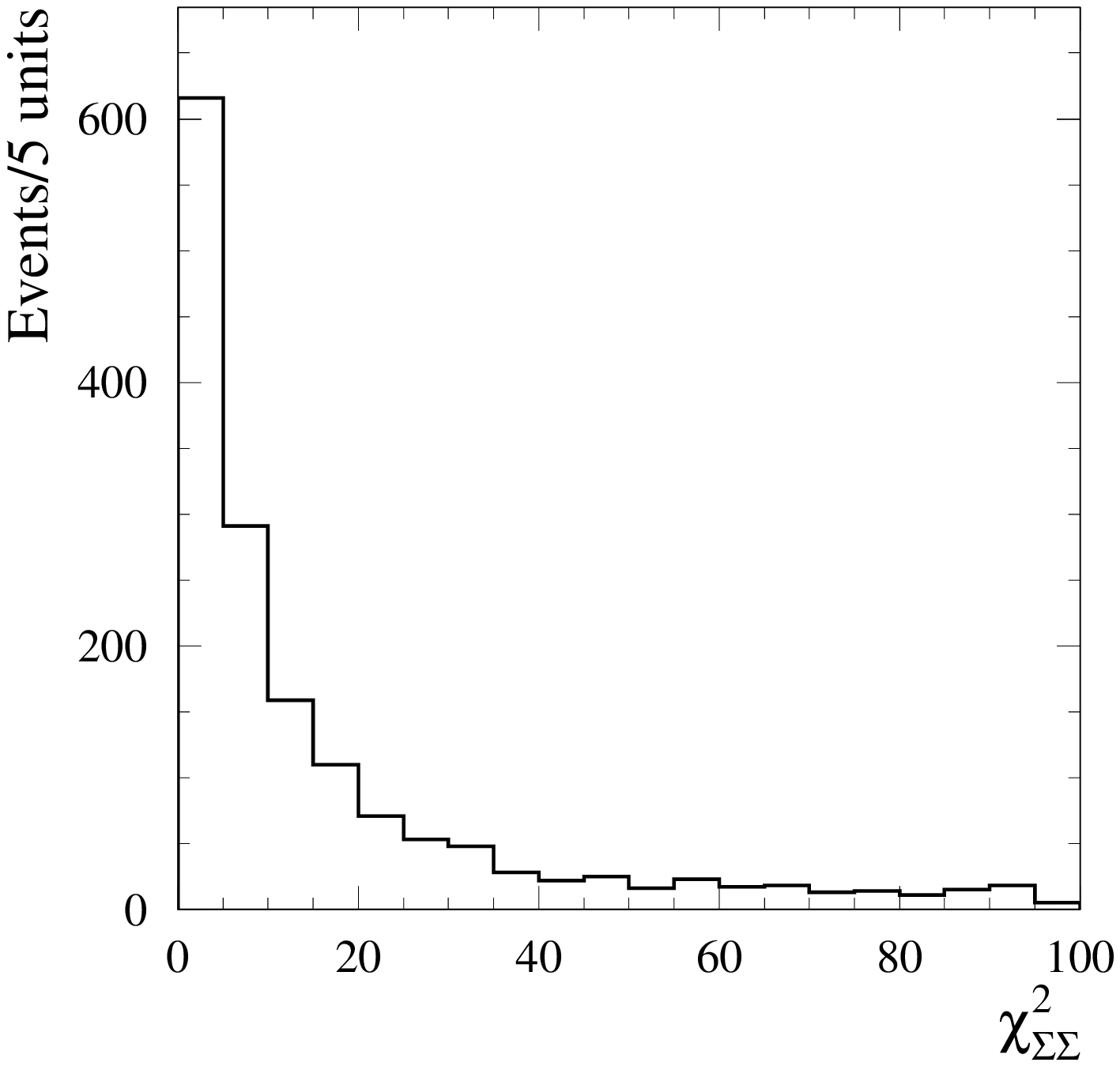}
\caption{The $\chi^2_{\Sigma\Sigma}$ distributions for simulated events 
for $e^+e^- \to \Sigma^0\Sigbar^0\gamma$.
\label{sigma_f0}}
\end{minipage}
\hspace{1mm}
\begin{minipage}[t]{0.32\textwidth}
\includegraphics[width=.98\textwidth]{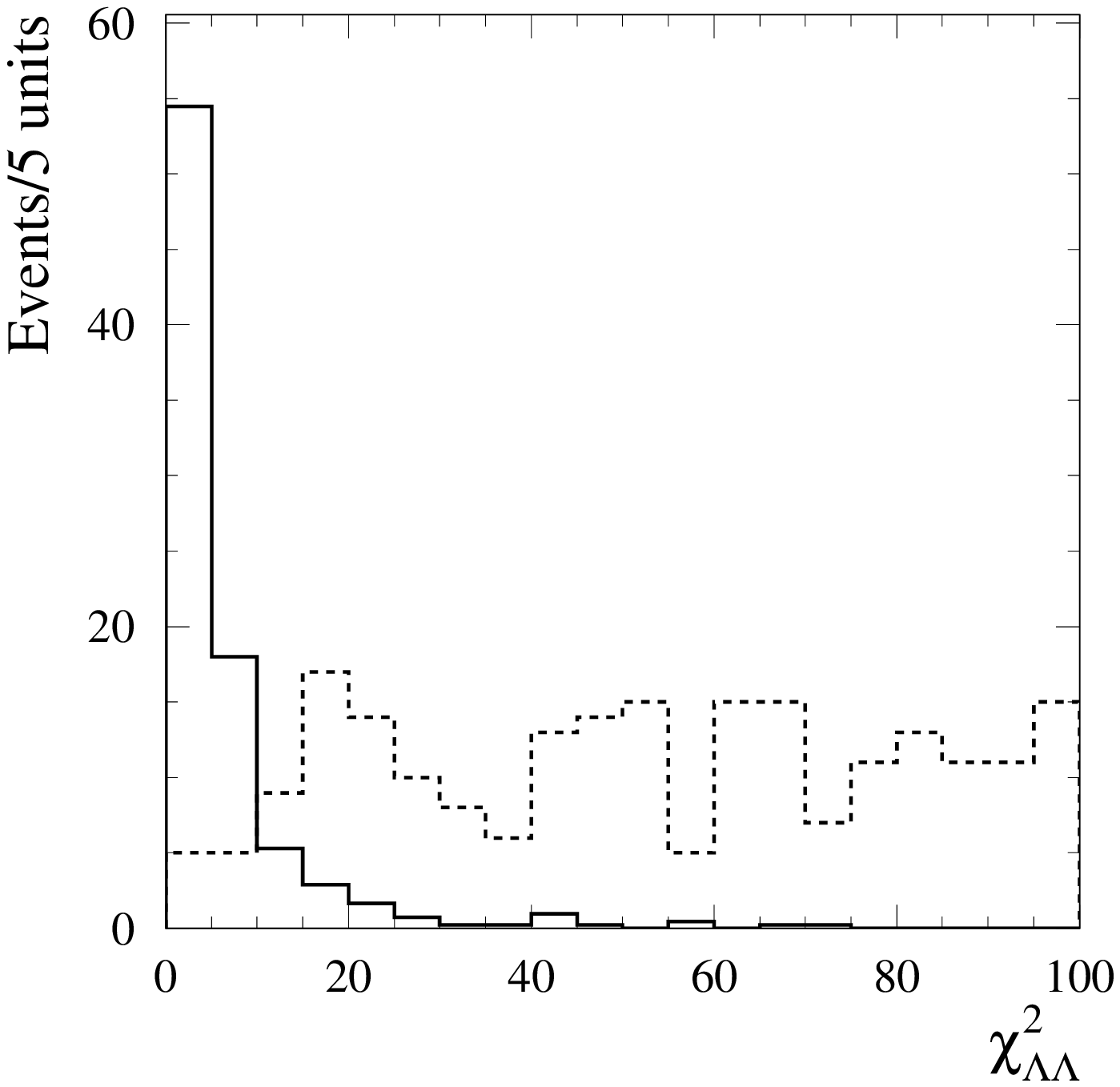}
\caption{The $\chi^2_{\Lambda\Lambda}$ distributions for simulated events 
for $e^+e^- \to \Lambda\Lbar\gamma$ (solid histogram) and
$e^+e^- \to \Sigma^0\Sigbar^0\gamma$ (dashed histogram).
\label{sigma_f1}}
\end{minipage}
\hspace{1mm}
\begin{minipage}[t]{0.32\textwidth}
\includegraphics[width=.98\textwidth]{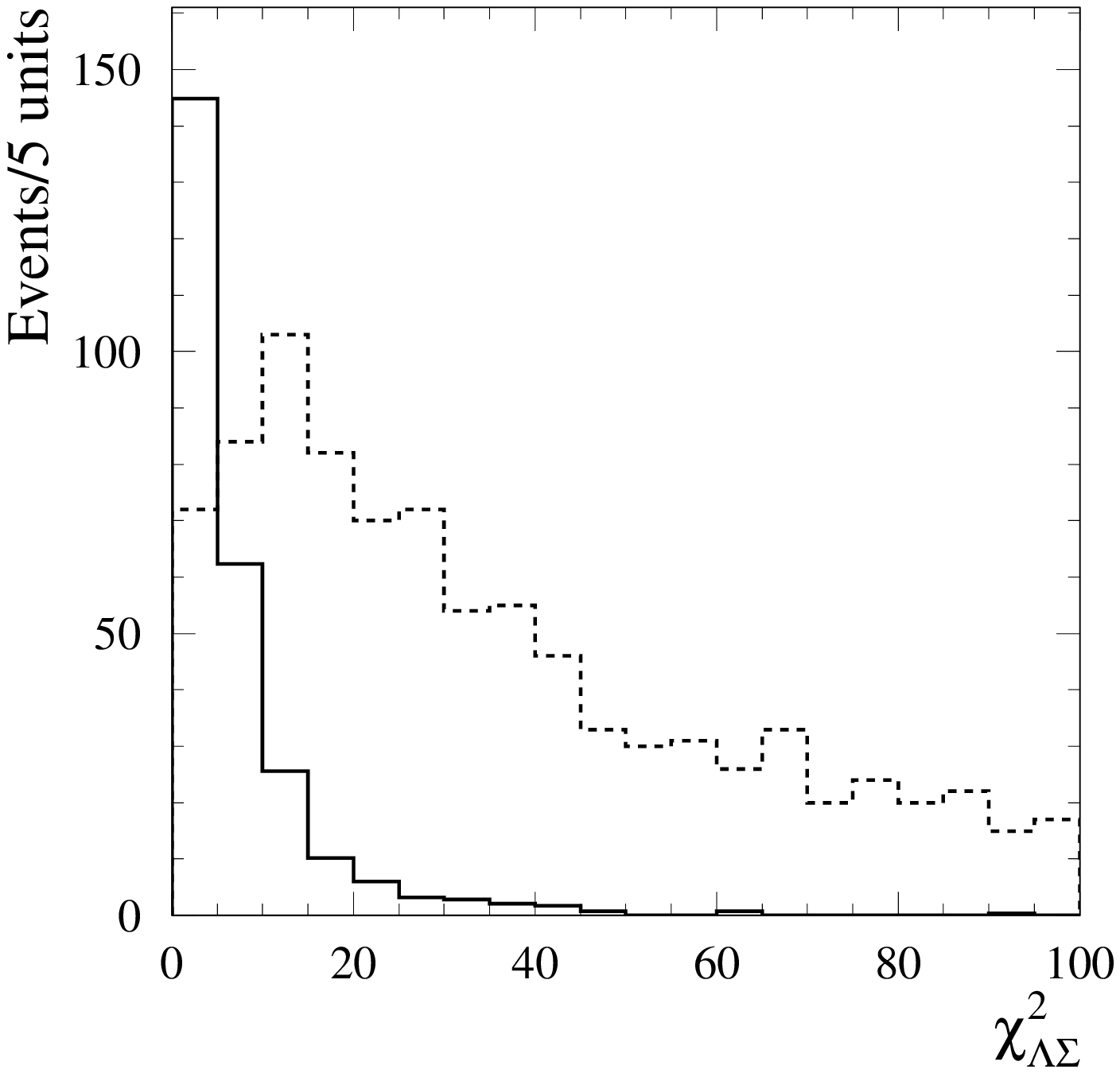}
\caption{The $\chi^2_{\Lambda\Sigma}$ distributions for simulated events 
for $e^+e^- \to \Lambda\Sigbar^0\gamma$ (solid histogram) and
$e^+e^- \to \Sigma^0\Sigbar^0\gamma$ (dashed histogram).
\label{sigma_f2}}
\end{minipage}
\end{figure*}

The distribution of the $\chi^2$ of the kinematic fit ($\chi^2_{\Sigma\Sigma}$) 
for simulated $e^+e^- \to \Sigma^0\Sigbar^0\gamma$ 
events is shown in Fig.\ref{sigma_f0}.
We select data events with $\chi^2_{\Sigma\Sigma}<20$ for further analysis;
as before, a $\chi^2$ control region ($20<\chi^2_{\Sigma\Sigma}<40$) is used
for background estimation and subtraction.

To suppress background from $e^+e^- \to \Lambda\Lbar\gamma$
and  $e^+e^- \to \Lambda\Sigbar^0\gamma$ events with extra
photons we also perform kinematic fits to the $\Lambda\Lbar\gamma$ 
and $\Lambda\Sigbar^0\gamma$ hypotheses.
The $\Lambda\Sigbar^0\gamma$ fit is a fit to the  
$e^+e^- \to \Lambda\Lbar\gamma\gamma$ hypothesis. 
The photon with highest $E^\ast_\gamma$ is assumed to be the ISR photon.
The other photon taken in combination with the $\Lambda$ or $\Lbar$,
must give an invariant mass value in the range 1.155--1.23 GeV/$c^2$,
where the  mass is calculated using fitted momenta. 
The $\chi^2_{\Lambda\Lambda}$ distributions for simulated events 
corresponding to $e^+e^- \to \Lambda\Lbar\gamma$ and 
$e^+e^- \to \Sigma^0\Sigbar^0\gamma$ are shown in 
Fig.\ref{sigma_f1}.
The requirement $\chi^2_{\Lambda\Lambda}>20$ rejects 93\% of
$\Lambda\Lbar\gamma$ events and only 3\% of signal events.
Similarly, the $\chi^2_{\Lambda\Sigma}$ distributions for simulated events 
for $e^+e^- \to \Lambda\Sigbar^0\gamma$ and 
$e^+e^- \to \Sigma^0\Sigbar^0\gamma$ are shown in 
Fig.\ref{sigma_f2}.
The requirement $\chi^2_{\Lambda\Sigma}>20$ again rejects 93\% of 
$\Lambda\Sigbar^0\gamma$ events, but in this case removes 30\% of 
the signal events.
Data events with $\chi^2_{\Lambda\Sigma}<20$ are used
to estimate the level of $\Lambda\Sigbar^0\gamma$ background.

\begin{figure*}
\includegraphics[width=.33\textwidth]{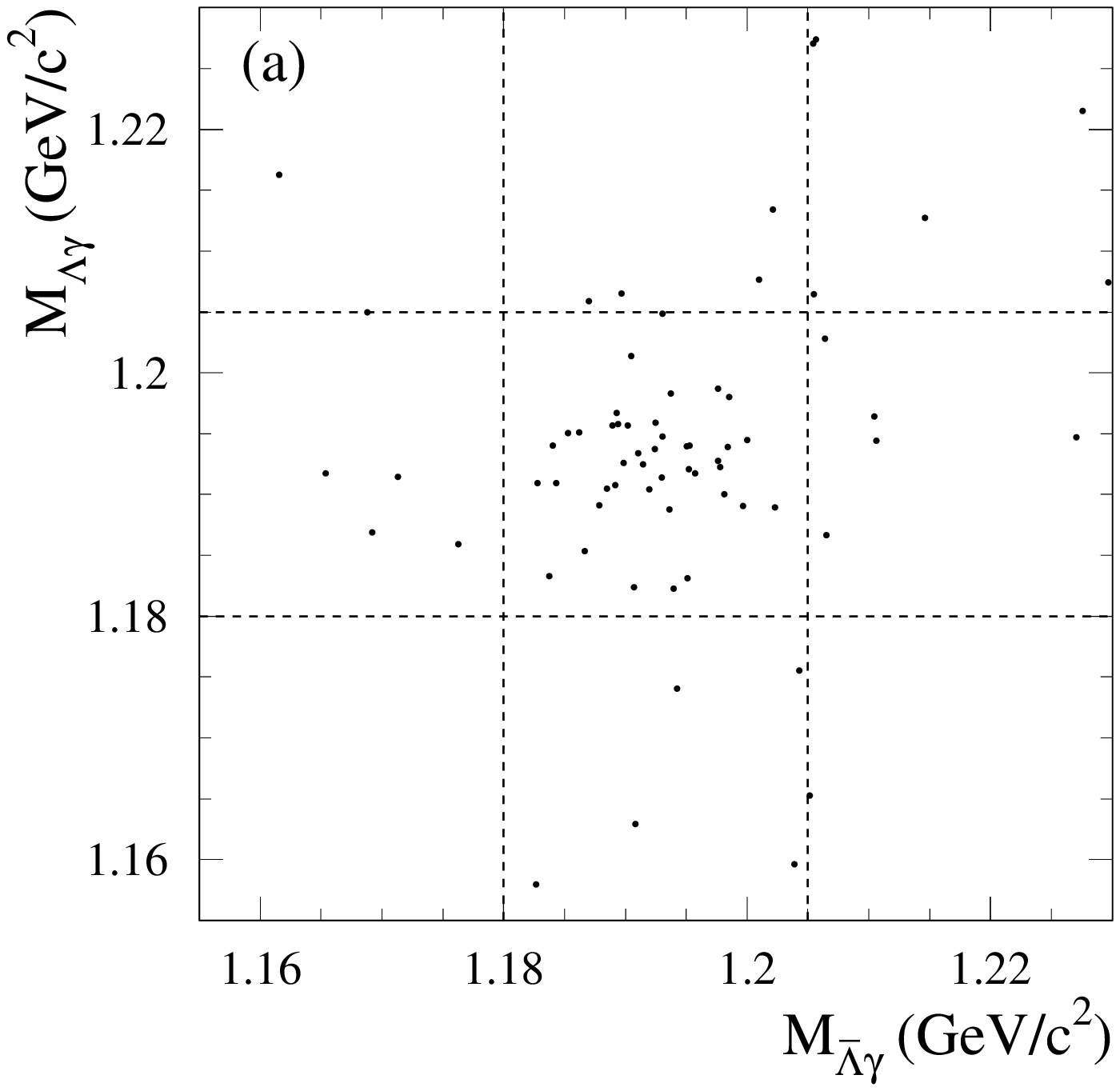}
\includegraphics[width=.33\textwidth]{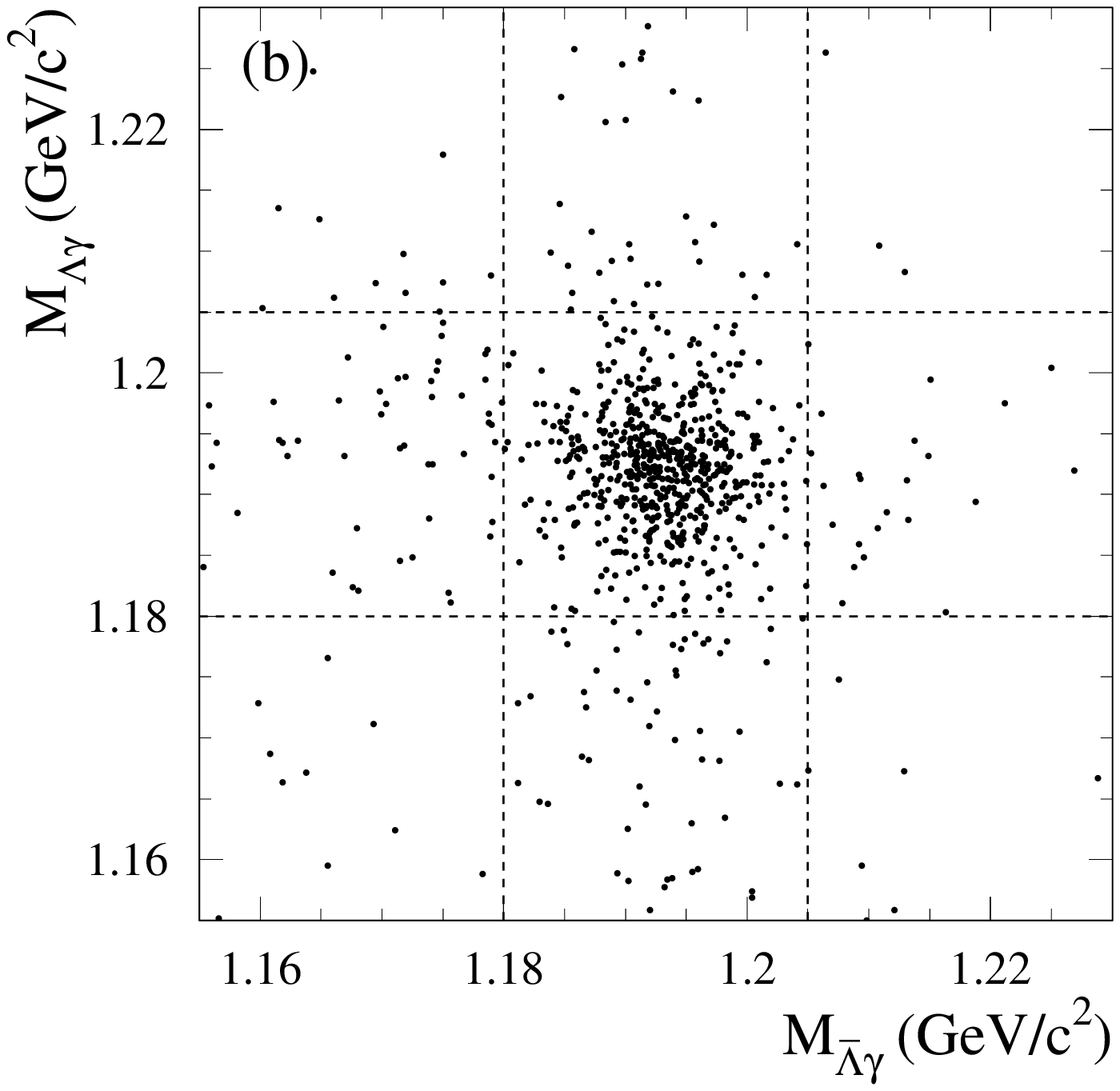}
\caption{Scatter plots of the invariant
mass of the $\Sigma^0$ candidate versus the invariant mass 
of the $\Sigbar^0$ candidate for selected data events (a), 
and simulated $e^+e^- \to \Sigma^0\Sigbar^0\gamma$ events
(b).
\label{sigma_f3}}
\end{figure*}

\begin{figure*}
\begin{minipage}[t]{0.33\textwidth}
\includegraphics[width=.98\textwidth]{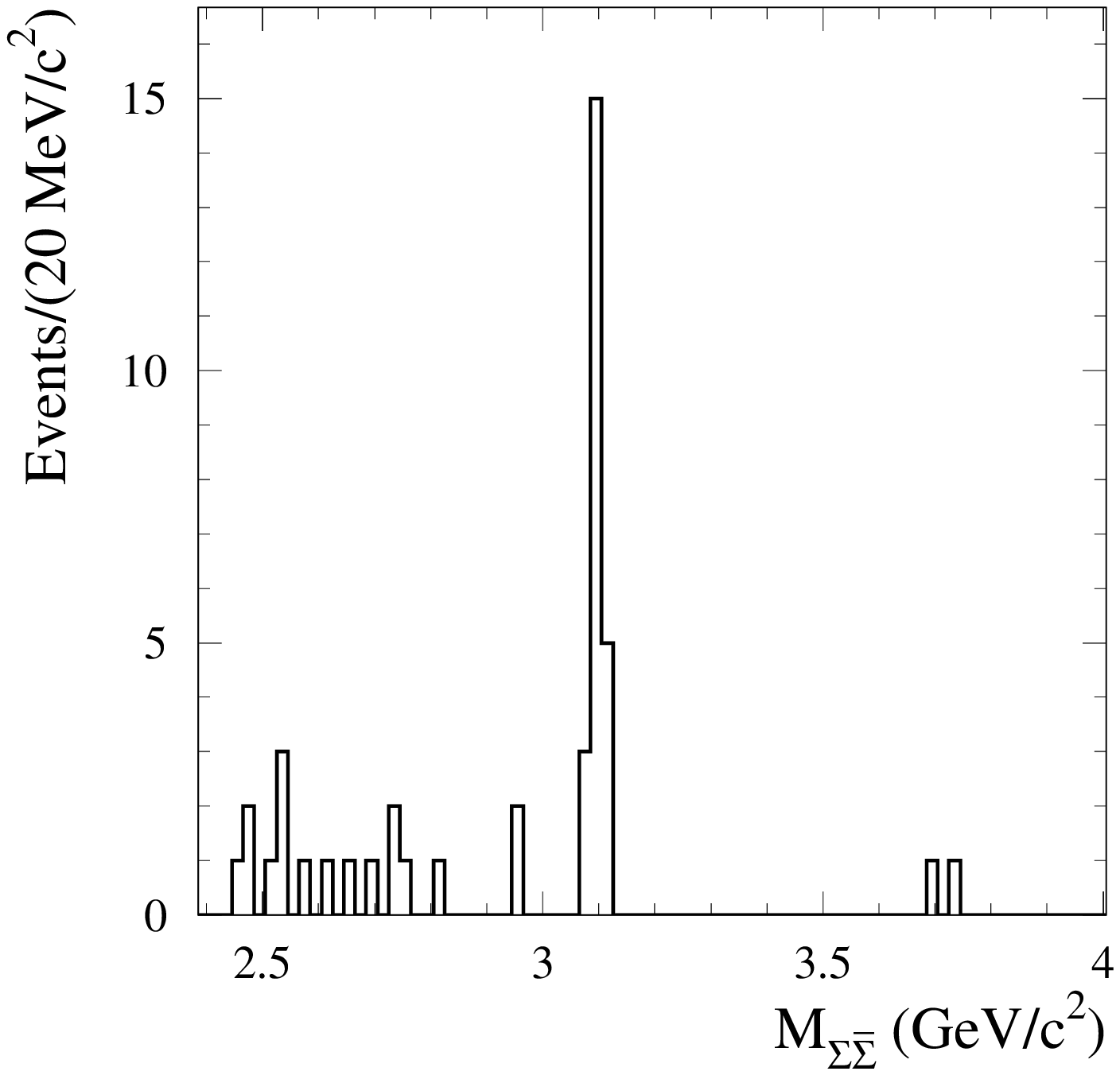}
\caption{The $\Sigma^0\Sigbar^0$ mass spectrum for
the selected data events.
\label{sigma_f4}}
\end{minipage}
\hspace{8mm}
\begin{minipage}[t]{0.33\textwidth}
\includegraphics[width=.98\textwidth]{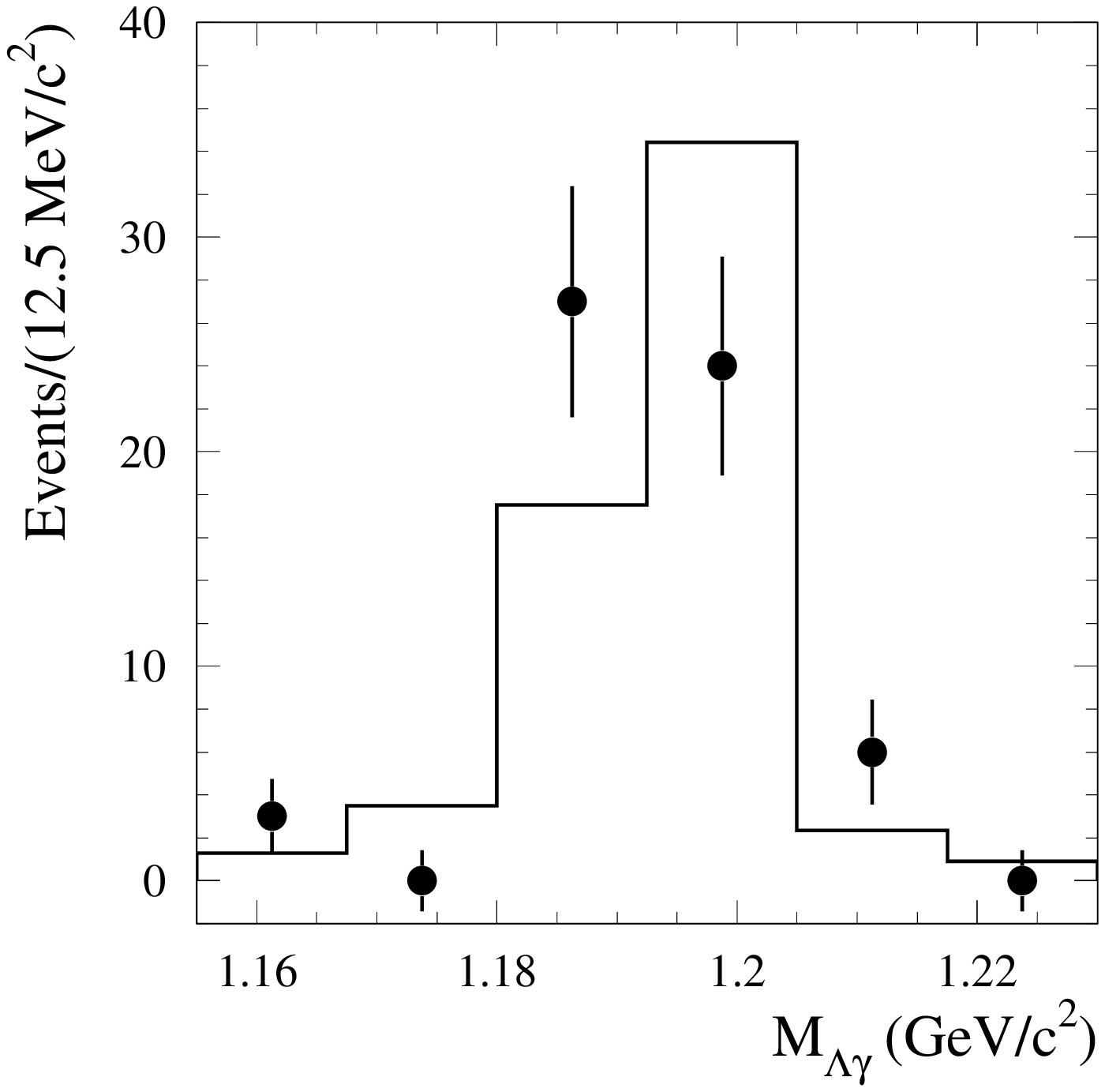}
\caption{The $\Lambda\gamma$ and $\Lbar\gamma$ invariant mass distribution 
for data (points with
error bars) and simulated (histogram) events from the $J/\psi$ region.
\label{sigma_f5}}
\end{minipage}
\end{figure*}

The scatter plots of the invariant mass of the $\Sigma^0$ candidate
versus the invariant mass of the $\Sigbar^0$ candidate for
the selected data events and
simulated $e^+e^- \to \Sigma^0\Sigbar^0\gamma$
events are shown in Figs.~\ref{sigma_f3}(a) and (b), respectively. 
Of the two possible $\Lambda\gamma$ and $\Lbar\gamma$ combinations, 
we plot only the combination with the smaller value of
$(M_{\Lambda\gamma}-m_\Sigma)^2+(M_{\Lbar\gamma}-m_\Sigma)^2$, where
$m_\Sigma$ is the nominal $\Sigma^0$ mass.
The $\Sigma^0\Sigbar^0$ mass spectrum for the data events 
with the additional requirement that the $\Sigma^0$ and $\Sigbar^0$ 
candidate mass values satisfy 
$1.180 < M_{\Lambda\gamma} < 1.205$ GeV/$c^2$ (central box in 
Fig.~\ref{sigma_f3}(a)), 
is shown in Fig.~\ref{sigma_f4}.
An excess of signal events is seen at masses below 3.0 GeV/$c^2$.
There are also about 20 events near the $J/\psi$ mass, corresponding to
$J/\psi\to\Sigma^0\Sigbar^0$ decay. The two events 
near 3.7 GeV/$c^2$ may be due to  $\psi(2S)\to\Sigma^0\Sigbar^0$
decay. The mass distribution for $\Sigma^0$ and $\Sigbar^0$candidates
from the $J/\psi$ region is shown in Fig.~\ref{sigma_f5}. The
spectrum is obtained as the difference of the spectrum from the region 
$3.05<M_{\Sigma\Sigbar}<3.15$ GeV/$c^2$ and that from the sideband region
(3.00--3.05 and 3.15--3.20 GeV/$c^2$). We see that simulation
reproduces the $\Sigma^0$ lineshape quite well.

\subsection{Background subtraction}
Background processes can be divided into 
three classes, namely those with zero 
($e^+e^-\to \Lambda\Lbar\gamma$, $\Lambda\Lbar\pi^0$,
$\Lambda\Lbar\pi^0\gamma$, {\ldots} ), 
one ($e^+e^-\to \Lambda\Sigbar^0\gamma$, $\Lambda\Sigbar^0\pi^0$,
$\Lambda\Sigbar^0\pi^0\gamma$, {\ldots} ),
and two $\Sigma^0$'s ($e^+e^-\to\Sigma^0\Sigbar^0\pi^0$,
$\Sigma^0\Sigbar^0\pi^0\gamma$, {\ldots} )
in the final state.
To separate events with two $\Sigma^0$'s from events with no $\Sigma^0$'s
and one $\Sigma^0$ we use the differences in their two-dimensional 
distributions of invariant mass values of the $\Sigma^0$ and $\Sigbar^0$ 
candidates. 

Background events from $e^+e^- \to \Sigma^0\Sigbar^0\pi^0$
with an undetected low-energy photon or with merged photons from $\pi^0$
decay yield a low value of $\chi^2$ when reconstructed under 
the $\Sigma^0\Sigbar^0\gamma$ hypothesis, and can not be separated          
from the process under study. Special selection procedures are
applied in order to estimate this background. The procedures are
similar to those used to study background from 
$e^+e^-\to\Lambda\Lbar\pi^0$ in Sec.\ref{background}. 
No $\Sigma^0\Sigbar^0\pi^0$ candidates are found in data, 
and we estimate that the background from this process does not 
exceed 5 events. Assuming that the dibaryon mass spectrum in
$e^+e^- \to \Sigma^0\Sigbar^0\pi^0$ is similar to that for
$e^+e^- \to \proton\antiproton\pi^0$~\cite{BADpp} we find that about 
70\% of the $e^+e^- \to \Sigma^0\Sigbar^0\pi^0$ events are
expected to have $\Sigma^0\Sigbar^0$ mass less than 3 GeV/$c^2$.
Two-$\Sigma^0$ background other than that from $\Sigma^0\Sigbar^0\pi^0$
can be estimated using the difference in the $\chi^2$ distributions 
for signal and background events.

The $3\times3$ two-dimensional histograms 
of $M_{\Lambda\gamma}$ vs $M_{\Lbar\gamma}$ (dashed lines in Fig.~\ref{sigma_f3}) 
for events from three classes:
\begin{eqnarray}
&\chi^2_{\Sigma\Sigma} < 20,\; \chi^2_{\Lambda\Sigma} > 20,\nonumber \\
&20 < \chi^2_{\Sigma\Sigma} < 40,\; \chi^2_{\Lambda\Sigma} > 20, \nonumber \\
&\chi^2_{\Sigma\Sigma} < 20,\; \chi^2_{\Lambda\Sigma} < 20, \nonumber
\end{eqnarray}
are fitted simultaneously. The second histogram is used to
determine two-$\Sigma^0$ background. From the third histogram we
estimate the $e^+e^-\to\Lambda\Sigbar^0\gamma$ background.
Each histogram is fitted using the following function:
$$N_{ij}=N_2 f^{2\Sigma}_{i,j}+N_1 f^{1\Sigma}_{i,j}+N_0 f^{0\Sigma}_{i,j},$$
where $N_0$, $N_1$, and $N_2$ are the numbers of events with zero, 
one, and two $\Sigma^0$'s in the final state. 
The functions $f^{1\Sigma}$ and $f^{2\Sigma}$ are taken from 
$e^+e^-\to \Lambda\Sigbar^0\gamma$ and
$e^+e^-\to \Sigma^0\Sigbar^0\gamma$ simulations. 
The probability density function for
zero-$\Sigma^0$ events is the product of two identical linear functions of 
$M_{\Lambda\gamma}$ and $M_{\Lbar\gamma}$ and a function taking into account 
the correlation between masses of the $\Sigma^0$ and $\Sigbar^0$ candidates. 
This last function is extracted from $e^+e^-\to \Lambda\Lbar\gamma$ 
simulation. The correlation arises from our choice of
one of the two possible combinations of $\Lambda$ and $\Lbar$ candidates 
with photons, and is about 15\% for the central mass bin. 

In order to find the number of signal events and estimate the background 
we use the following relations:
\begin{eqnarray}
&N_2(\chi^2_{\Sigma\Sigma} < 20,\; \chi^2_{\Lambda\Sigma} > 20)= 
N_{2s}+N_{2b}, \nonumber \\
&N_2(20<\chi^2_{\Sigma\Sigma} < 40,\; \chi^2_{\Lambda\Sigma} > 20)= 
\alpha_1 N_{2s}+\beta_1 N_{2b}. \nonumber \\
&N_2(\chi^2_{\Sigma\Sigma} < 20,\; \chi^2_{\Lambda\Sigma} < 20)= 
\alpha_2 N_{2s}+\beta_2 N_{2b}, \nonumber \\
&N_1(\chi^2_{\Sigma\Sigma} < 20,\; \chi^2_{\Lambda\Sigma} > 20)= 
N_{1s}+N_{1b}, \nonumber \\
&N_1(20<\chi^2_{\Sigma\Sigma} < 40,\; \chi^2_{\Lambda\Sigma} > 20)= 
\gamma_1 N_{1s}+\delta_1 N_{1b}. \nonumber \\
&N_1(\chi^2_{\Sigma\Sigma} < 20,\; \chi^2_{\Lambda\Sigma} < 20)= 
\gamma_2 N_{1s}+\delta_2 N_{1b}, \nonumber 
\end{eqnarray}
where $N_{2s}$ is the number of signal $\Sigma^0\Sigbar^0\gamma$
events and $N_{2b}$ is the number of two-$\Sigma^0$ background events in the
signal region ($\chi^2_{\Sigma\Sigma} < 20,\; \chi^2_{\Lambda\Sigma} > 20$),
$N_{1s}$ is the number of $\Lambda\Sigbar^0\gamma$ events and 
$N_{1b}$ is the number of one-$\Sigma^0$ events from all other processes in
the signal region; $N_{2s}$, $N_{2b}$, $N_{1s}$, 
and $N_{1b}$, are then free fit parameters. 
The coefficients $\alpha_i$, $\beta_i$, $\gamma_i$,
and $\delta_i$ are obtained from simulation of 
$e^+e^-\to \Sigma^0\Sigbar^0\gamma$,
$e^+e^-\to \Sigma^0\Sigbar^0\pi^0\gamma$, 
$e^+e^-\to\Lambda\Sigbar^0\gamma$,
$e^+e^-\to\Lambda\Sigbar^0\pi^0\gamma$, respectively.
For the coefficients most critical to the analysis,
$\alpha_1=0.22\pm0.03$ and $\beta_1=1.5\pm0.3$,
the errors include uncertainties due to the data-MC difference in
the $\chi^2$ distributions for the  kinematic fits.
The other 6 free parameters are the numbers of zero-$\Sigma^0$ events
in the three histograms, and the slopes of the linear functions
describing the mass distributions for these events.
\begin{table*}
\caption{Comparison of the fit results for $\Sigma^0\Sigbar^0$ masses below 3 GeV/$c^2$ and
the predictions from JETSET simulation;
$N_{2s}$, $N_0$, $N_1$, $N_{2b}$ are the fitted numbers
of signal, zero-, one-, and two-$\Sigma^0$ background events in the signal
region ($\chi^2_{\Sigma\Sigma} < 20,\; \chi^2_{\Lambda\Sigma} > 20$), respectively,
$N_{\Sigma^0\Sigbar^0\pi^0}$ is expected number of background events
from $e^+e^-\to \Sigma^0\Sigbar^0\pi^0$ process.
\label{nsigsig}\\}
\begin{ruledtabular}
\begin{tabular}{lccccc}
      &        $N_{2s}$        & $N_{0}$      & $N_{1}$     & $N_{2b}$    & $N_{\Sigma^0\Sigbar^0\pi^0}$\\
data  &  $18.1^{+7.8}_{-7.5}$  & $11.3\pm4.8$ & $1.2\pm0.6$ & $2.8\pm4.8$ &    $< 4$ \\   
JETSET&         $33\pm5$       &  $3.1\pm1.4$ & $1.2\pm0.8$ &   $ < 1.4$  &   $0.6\pm0.6$ \\
\end{tabular}
\end{ruledtabular}
\end{table*}

The fit results for $\Sigma^0\Sigbar^0$ masses below 3 GeV/$c^2$ are
shown in Table~\ref{nsigsig}, together with the predictions from JETSET 
simulation. The one-$\Sigma^0$ background is dominated by the 
$e^+e^-\to\Lambda\Sigbar^0\gamma$ process. The number of one-$\Sigma^0$ events
from other processes is found to be consistent with zero. The numbers of 
$e^+e^-\to\Sigma^0\Sigbar^0\gamma$ and $e^+e^-\to\Lambda\Sigbar^0\gamma$ events
with $\chi^2_{\Lambda\Sigma} < 20$ are $7.7^{+3.4}_{-3.2}$ 
and $15.3^{+5.4}_{-7.7}$, respectively.

The fitting procedure was performed in five $\Sigma^0\Sigbar^0$ mass 
ranges, and the number of signal events found in each is listed 
in Table~\ref{sigsigt}.
For $\Sigma^0\Sigbar^0$ masses below 3 GeV/$c^2$ we observe an excess of 
signal events over background. The significance of the observation of 
$\Sigma^0\Sigbar^0$ production in the mass region below 
3.0 GeV/$c^2$ is 2.9$\sigma$. For other mass bins we list upper limits
at the 90\% CL. 

\subsection{Cross section and form factor}\label{crosssec1}
     The cross section for $e^+e^-\to \Sigma^0\Sigbar^0$ is calculated from
the $\Sigma^0\Sigbar^0$ mass spectrum according to 
Eqs.(\ref{ISRcs}-\ref{ISRlum}).

The detection efficiency is determined from MC simulation and then
corrected for data-MC simulation differences in detector response.
The model dependence of the detection efficiency due to 
the unknown $|G_E/G_M|$ ratio is estimated to be 5\% (see Sec.~\ref{efflam}).
The efficiency corrections summarized in Table~\ref{tab_ef1_cor} were
discussed in Sec.~\ref{efflam}.
\begin{table}
\caption{The values of the various efficiency corrections
for the process $e^+e^-\to \Sigma^0\Sigbar^0\gamma$.
\label{tab_ef1_cor}\\}                                
\begin{ruledtabular}
\begin{tabular}{ll}
effect                &$\delta_i$, (\%) \\
\hline
$\chi^2_{\Sigma\Sigma} < 20$         & $-2.0\pm6.0$  \\
track reconstruction                 & $-1.0\pm3.8$  \\
$\antiproton$ nuclear interaction        & $+1.0\pm0.4$  \\
PID                                  & $+0.6\pm0.6$  \\
photon inefficiency                  & $-3.9\pm0.9$  \\
photon conversion                    & $+1.2\pm0.6$  \\
\hline 
total                                & $-4.1\pm7.2$ \\
\end{tabular} 
\end{ruledtabular}
\end{table}        
On the basis of our analysis of ISR processes with photons
in the final state~\cite{BAD5pi}, we enlarge the systematic error 
in the correction for the $\chi^2$ selection interval. The correction 
for trigger inefficiency is removed, since for
$e^+e^-\to \Sigma^0\Sigbar^0\gamma$, trigger
inefficiency is less than 0.001 both in data and in MC simulation.
The corrected detection efficiencies are listed in Table~\ref{sigsigt}.
The overall uncertainty in efficiency takes into account simulation 
statistical error, model uncertainty, the error in the $\Lambda \to \proton\pi^-$
branching fraction, and the uncertainty of the efficiency correction. 
\begin{table*}
\caption{The $\Sigma^0\Sigbar^0$ invariant mass interval ($M_{\Sigma\Sigbar}$),
net number of signal events ($N_s$),
detection efficiency ($\varepsilon$), ISR luminosity ($L$),
measured cross section ($\sigma$), and effective form factor ($F$) 
for $e^+e^-\to \Sigma^0\Sigbar^0$. The quoted errors on $\sigma$
are statistical and systematic. For the form factor, the total error
is listed.
\label{sigsigt}\\}
\begin{ruledtabular}
\begin{tabular}{ccccccc}
$M_{\Sigma\Sigbar}$ & $N_s$ & $\varepsilon$ &   $L$    & $\sigma$ & $|F|$ \\
(GeV/$c^2$)    &       &               & (pb$^{-1}$) &  (pb)    &       \\
\hline\\
2.385--2.600&$ 10.3_{-4.5}^{+4.4}$  &$0.024\pm0.002$&  14.3&$ 30\pm13\pm3$ &$0.090^{+0.018}_{-0.023}$\\
2.600--2.800&$  6.5_{-3.8}^{+3.1}$  &$0.025\pm0.003$&  14.7&$ 17^{+8}_{-10}\pm 2$ &$0.047^{+0.010}_{-0.017}$\\
2.800--3.000&$  1.4_{-3.2}^{+3.5}$  &$0.026\pm0.003$&  16.1&$ 3.4^{+8.5}_{-7.8}\pm 0.4 $&$ 0.021^{+0.018}_{-0.021}$                \\
3.200--3.600&$ < 2.3$               &$0.023\pm0.003$&  39.9&$   < 2.5$            &$< 0.019$                \\
3.800--5.000&$ < 2.3$               &$0.023\pm0.002$& 180.4&$   < 0.5$            &$< 0.011$                \\
\end{tabular}
\end{ruledtabular}
\end{table*}

The measured values of the $e^+e^-\to \Sigma^0\Sigbar^0$
cross section are listed in Table~\ref{sigsigt}, together with
the effective form factor values calculated according to 
Eq.(\ref{efform}). 
The quoted errors on the cross section are statistical and
systematic. The latter includes systematic uncertainty in
detection efficiency, the uncertainty in total integrated luminosity
(1\%), and radiative correction uncertainty (1\%).
This is the first measurement of the $e^+e^-\to \Sigma^0\Sigbar^0$
cross section. The upper limit set by DM2~\cite{DM2ll} at
2.386 GeV ( $< 120$ pb) is consistent with our measurements.
\begin{figure}
\includegraphics[width=.33\textwidth]{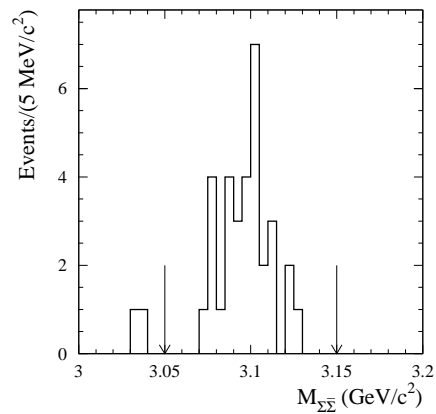}
\caption{The $\Sigma^0\Sigbar^0$ mass spectrum 
for the mass region near the $J/\psi$. The arrows
indicate the boundaries between the signal region 
and sideband regions.
\label{psiss}}
\end{figure}
\subsection{\boldmath $J/\psi$ decay into $\Sigma^0\Sigbar^0$}\label{jpsi1}
The $\Sigma^0\Sigbar^0$ mass spectrum for selected events in the 
$J/\psi$ mass region is shown in Fig.~\ref{psiss}.
We determine the number of resonance events by counting the events in
the signal region indicated in Fig.~\ref{psiss} and subtracting the
number in the two sidebands.
The net number of $J/\psi$ decay events is then $30\pm 6$.

The detection efficiency is estimated from MC simulation.
The event generator uses the experimental data on the $\Sigma^0$
angular distribution in $J/\psi \to \Sigma^0\Sigbar^0$ decay,
which is $1+\alpha \cos^2theta$
with $\alpha=-0.1\pm0.2$~\cite{MARKll,DM2psi,BESll}.
The error in the detection efficiency due to 
the uncertainty in $\alpha$ is negligible.
The efficiency corrected for data-MC simulation differences is
$\varepsilon$ = 0.022$\pm$0.002.

Using Eqs.~(\ref{eqpsi},\ref{eqpsi1}) the following product is         
determined:         
$$\Gamma(J/\psi\to e^+e^-){\cal B}(J/\psi\to \Sigma^0\Sigbar^0)=
(6.4\pm 1.2\pm 0.6)\mbox{ eV}.$$ 
The systematic error includes the uncertainties in detection efficiency,
integrated luminosity, and in the radiative correction.         
Using the PDG value of the electronic width~\cite{pdg},
the $J/\psi\to \Sigma^0\Sigbar^0$ branching         
fraction is calculated to be         
$${\cal B}(J/\psi\to \Sigma^0\Sigbar^0)=(1.16\pm0.26)\times 10^{-3}.$$
Our result is in agreement with the world average value
$(1.31\pm0.10)\times 10^{-3}$~\cite{pdg}.

We also observe 2 events in the $\psi(2S)$ region with zero background,
estimated from the sidebands. This number agrees with the $2.5\pm0.4$
events expected from the measured branching fraction 
${\cal B}(\psi(2S)\to \Sigma^0\Sigbar^0)=
(2.51\pm0.31)\times 10^{-4}$~\cite{cleo2s,bes2s}.

\section{\boldmath The reaction $e^+e^- \to \Lambda\Sigbar^0\gamma$}
\subsection{Event selection}
The preliminary selection of  
$e^+e^- \to \Lambda\Sigbar^0\gamma$
events is similar to that for $e^+e^- \to \Lambda\Lbar\gamma$.
Additionally we require that an event candidate contain at least one extra 
photon with energy  greater than 30 MeV.
To suppress combinatorial background from events not containing 
two $\Lambda$'s in the final state we require that
the mass of the $\Lambda$ ($\Lbar$) satisfy 
$1.110 < M_{\proton\pi^-} < 1.122$ GeV/$c^2$.

\begin{figure}
\includegraphics[width=.33\textwidth]{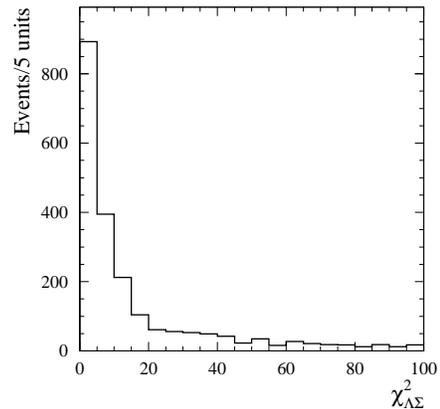}
\caption{The $\chi^2_{\Lambda\Sigma}$ distributions for simulated 
$e^+e^- \to \Lambda\Sigbar^0\gamma$ events.
\label{lamsig_f1}}
\end{figure}

For events passing the preliminary selection, we perform a kinematic fit to
the $e^+e^- \to \Lambda\Sigbar^0\gamma$ hypothesis 
as described in Sec.~\ref{sigmasel}.
The $\chi^2_{\Lambda\Sigma}$ distribution for simulated 
$\Lambda\Sigbar^0\gamma$ events is shown in Fig.~\ref{lamsig_f1}.
We select the events with $\chi^2_{\Lambda\Sigma}<20$
for further analysis. The control region ($20<\chi^2_{\Lambda\Sigma}<40$)
is used for background estimation and subtraction.
\begin{figure}
\includegraphics[width=.33\textwidth]{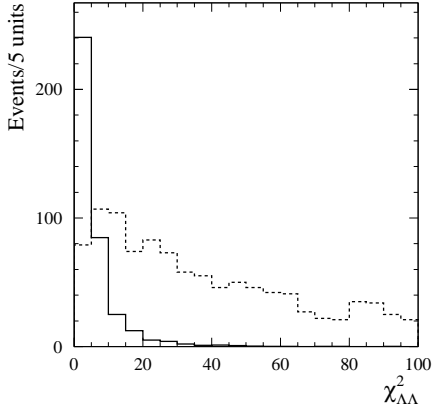}
\caption{The $\chi^2_{\Lambda\Lambda}$ distributions for simulated 
$e^+e^- \to \Lambda\Lbar\gamma$ events (solid histogram) and
$e^+e^- \to \Lambda\Sigbar^0\gamma$ events(dashed histogram).
\label{lamsig_f2}}
\end{figure}
To suppress background resulting from $e^+e^- \to \Lambda\Lbar\gamma$
events with an additional photon we perform a kinematic fit to 
the $\Lambda\Lbar\gamma$ hypothesis and 
require that $\chi^2_{\Lambda\Lambda}>20$.
The $\chi^2_{\Lambda\Lambda}$ distributions for simulated 
$e^+e^- \to \Lambda\Lbar\gamma$ and 
$e^+e^- \to \Lambda\Sigbar^0\gamma$ events are shown in 
Fig.~\ref{lamsig_f2}.
The $\chi^2_{\Lambda\Lambda}>20$ cut rejects 95\% of 
the $\Lambda\Lbar\gamma$ events at the cost of 20\% of 
the signal events.

\begin{figure}
\includegraphics[width=0.33\textwidth]{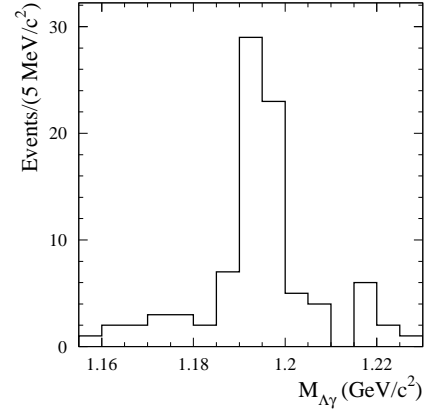}
\caption{The distribution of the invariant mass of the $\Sigma^0$
and $\Sigbar^0$ candidates for the selected $\Lambda\Sigbar^0\gamma$
candidate.
\label{lamsig_f3}}
\end{figure}

\begin{figure}
\includegraphics[width=0.33\textwidth]{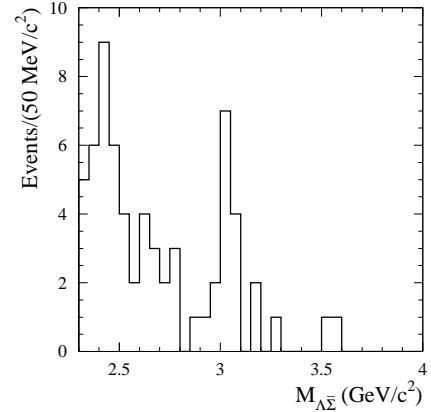}
\caption{The $\Lambda\Sigbar^0$ invariant mass spectrum for data events
with the invariant mass of the $\Sigma^0$ candidate in the 1.185-1.205 
GeV/$c^2$ range.
\label{lamsig_f4}}
\end{figure}

The distribution of $\Sigma^0$
candidate invariant mass
for data events passing the $\Lambda\Sigbar^0\gamma$ selection
process is shown in Fig.~\ref{lamsig_f3}.
For each event we plot only the $\Lambda(\Lbar)\gamma$ combination
closer to the nominal $\Sigma^0$ mass. 
The $\Lambda\Sigbar^0$ mass distribution for selected
data events with invariant mass of
the $\Sigbar^0$ candidate in the 1.185-1.205 GeV/$c^2$ range is
shown in Fig.~\ref{lamsig_f4}. We expect that
the $e^+e^- \to \Sigma^0\Sigbar^0\gamma$ process 
results in a significant contribution to the selected event sample. 
In particular, 
the peak in the $\Lambda\Sigbar^0$ mass spectrum near 3 GeV/$c^2$
is from $J/\psi\to \Sigma^0\Sigbar^0$ events with a missing or
excluded photon.

\subsection{Background subtraction} 
The background processes can be divided into 
three classes, namely those with zero 
($e^+e^-\to \Lambda\Lbar\gamma$, $\Lambda\Lbar\pi^0$,
$\Lambda\Lbar\pi^0\gamma$, {\ldots} ), 
one ($e^+e^-\to \Lambda\Sigbar^0\pi^0$, $\Lambda\Sigbar^0\pi^0\gamma$, {\ldots} ),
and two $\Sigma^0$'s ($e^+e^-\to\Sigma^0\Sigbar^0\gamma$,
$\Sigma^0\Sigbar^0\pi^0$, $\Sigma^0\Sigbar^0\pi^0\gamma$, {\ldots} )
in the final state.
To separate one-$\Sigma^0$ events from events with no  
$\Sigma^0$ we use the difference in mass  
distribution for the respective $\Sigma^0$ ($\Sigbar^0$) candidates. 

To determine two-$\Sigma^0$ background we select
a clean sample of two-$\Sigma^0$ events. 
To do this the $\Sigma^0\Sigbar^0\gamma$ criteria 
(Sec.~\ref{sigmasel}) are used with the additional requirements that
$1.180 <M_{\Lambda\gamma}  < 1.205$ GeV/$c^2$  for the
$\Sigma^0$ and $\Sigbar^0$ candidates, and that $\chi^2_{\Lambda\Sigma} < 100$.
The latter requirement is needed to obtain a useful $M_{\Lambda\Sigbar}$
distribution.
Using the ratio of detection efficiencies for two-$\Sigma^0$ and 
$\Lambda\Sigbar^0\gamma$ selections ($\kappa=0.80\pm0.05$),
we can convert the number of events in the two-$\Sigma^0$ sample 
to an estimate of the number of background events in the 
$\Lambda\Sigbar^0\gamma$ sample.

The background from $e^+e^- \to \Lambda\Sigbar^0\pi^0$ events
with an undetected low-energy photon or with merged photons from $\pi^0$
decay cannot be separated from the process under study. 
The experimental data with special
selection criteria are used to estimate this background. The procedure is
similar to that used in the study of $e^+e^-\to\Lambda\Lbar\pi^0$ 
background in Sec.\ref{background}. We selected two $\Sigma^0\Sigbar^0\pi^0$
candidates with an expected background from $e^+e^-\to \Lambda\Lbar\gamma$ and
$e^+e^-\to \Lambda\Sigbar^0\gamma$ processes of $0.5\pm0.2$ events.
To suppress the $\Lambda\Sigbar^0\pi^0$ background in the 
$\Lambda\Sigbar^0\gamma$ sample we reject events with
$0.10<M_{2\gamma}<0.17$ GeV/$c^2$, where $M_{2\gamma}$ is
the invariant mass of the most energetic photon in an event and
another photon with energy greater than 0.1 GeV. This removes
about 1/3 of $\Lambda\Sigbar^0\pi^0$ events and less than 1\% of
signal events. After applying this selection criterion
the rates at which $e^+e^- \to \Lambda\Sigbar^0\pi^0$ events
are selected as $\Lambda\Sigbar^0\gamma$ or $\Lambda\Sigbar^0\pi^0$
are in the ratio $(2.1\pm0.2)$, and
the $\Lambda\Sigbar^0\pi^0$ background in the $\Lambda\Sigbar^0\gamma$ 
event sample is estimated to be $(3.1\pm2.2)$ events.
We assume that the dibaryon mass distribution for
the $e^+e^- \to \Lambda\Sigbar^0\pi^0$ process is similar
to that for the $e^+e^- \to \proton\antiproton\pi^0$ process~\cite{BADpp}.
In particular, about 70\% of the $e^+e^- \to \Lambda\Sigbar^0\pi^0$
events have $\Lambda\Sigbar^0$ mass less than 2.9 GeV/$c^2$.
Both observed $\Lambda\Sigbar^0\pi^0$ events lie in this mass region.

The one-$\Sigma^0$ background other than $\Lambda\Sigbar^0\pi^0$
is estimated using the difference in the $\chi^2$ distributions for 
signal and background events. Two histograms of $M_\Lambda\gamma$ 
for events with $\chi^2_{\Lambda\Sigma} < 20$ and
with $20 < \chi^2_{\Lambda\Sigma} < 40$ are fitted
simultaneously to the sum of the distributions for signal and background
\begin{equation}
n_i=N_{1}H_{1i}+N_{2}H_{2i}+N_0H_{0i},
\end{equation}
where $N_{1}$, $N_{2}$, and $N_0$ are the
numbers of events containing one, two, and zero $\Sigma^0$'s in the final 
state, respectively. The one-$\Sigma^0$ events are the signal events with
a possible contribution from the background processes 
$\Lambda\Sigbar^0\pi^0$, $\Lambda\Sigbar^0\pi^0\gamma$, etc.
The function $H_{1}$ describing the mass distribution
of one-$\Sigma^0$ events is calculated using 
$e^+e^-\to\Lambda\Sigbar^0\gamma$ simulation.
The distribution of two-$\Sigma^0$ events is taken from
$e^+e^-\to\Sigma^0\Sigbar^0\gamma$ simulation.     
The parameter $N_{2}$ is fixed by addition of the term 
$-\ln{f_P(n_0;\mu_0)}$ to minimize the likelihood function.
Here $f_P$ is a Poisson distribution, 
$n_0$ is number of events in the two-$\Sigma^0$ sample described above and
$\mu_0=\kappa r N_{2}$. The scale factor $\kappa=0.80\pm0.05$  is
found from $e^+e^-\to\Sigma^0\Sigbar^0\gamma$ simulation as the
ratio of detection efficiencies for two-$\Sigma^0$ and
$\Lambda\Sigbar^0\gamma$ selections.
The factor $r\approx1.1$ takes into account the purity of the two-$\Sigma^0$ 
sample, which is $(90\pm5)\%$. It should be noted that the two-$\Sigma^0$ 
sample contains not only $\Sigma^0\Sigbar^0\gamma$ but also events
from the processes $e^+e^-\to \Sigma^0\Sigbar^0\pi^0$, 
$\Sigma^0\Sigbar^0\pi^0\gamma$, etc. A similar approach is used
to introduce the $\Lambda\Sigbar^0\pi^0$ background into the fit.
The shape of the zero-$\Sigma^0$ background ($H_0$) is modeled 
using the mass distribution for the $e^+e^- \to \Lambda\Lbar\gamma$ process.
The distribution is parametrized as 
$H_{0}=(1+a_0(M_{\Lambda\gamma}-m_\Sigma))f(M_\Sigma)$, 
where $m_\Sigma$ is
nominal $\Sigma^0$ mass. The function $f(M_{\Lambda\gamma})$ 
describes the
deviation from a linear function due to our choice of one of
the two $\Lambda(\Lbar)\gamma$ combinations. This function is equal
to unity at the end points of the mass interval, and is about 2
at the center. We checked that the function $H_{0}$ with $a_0$ as free
parameter provides a good description of the mass distributions for 
simulated $e^+e^- \to \Lambda\Lbar\pi^0$ and 
$e^+e^- \to \Lambda\Lbar\pi^0\gamma$ events, 
and data $\Lambda\Lbar\gamma$ events selected by requiring
$\chi^2_{\Lambda\Lambda} <20$ and $\chi^2_{\Lambda\Sigma} <20$.

The one-$\Sigma^0$ background other than that from $\Lambda\Sigbar^0\pi^0$ 
is estimated from the fit according to Eq.(\ref{2bsub}).
The coefficients $\beta_{sig}$ and $\beta_{bkg}$ are
obtained from the signal and $\Lambda\Sigbar^0\pi^0\gamma$
simulation and take the values $0.15\pm0.02$ and $1.5\pm0.3$,
respectively. Their errors are enlarged to take into account
the data-MC simulation difference in the $\chi^2$ distributions
resulting from the kinematic fits.

\begin{figure}
\includegraphics[width=.33\textwidth]{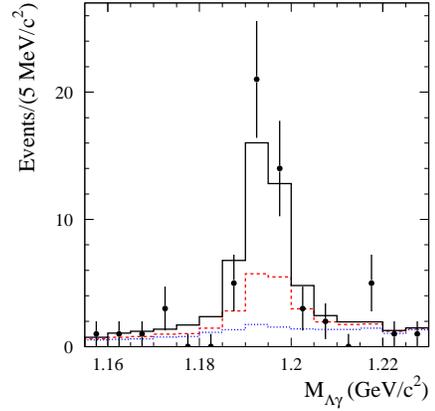}
\caption{
The distribution of the invariant mass of the $\Sigma^0$
($\Sigbar^0$) candidate for data events with 
$M_{\Lambda\Sigbar} < 2.9$ GeV/$c^2$
(points with error bars).              
The solid histogram shows the result of the fit described in the text.                   
The dotted histogram shows the contribution of zero-$\Sigma^0$
background. The difference between dashed and dotted histograms is
the contribution of two-$\Sigma^0$ background. 
\label{lamsig_f5}}
\end{figure}
\begin{figure}
\includegraphics[width=.33\textwidth]{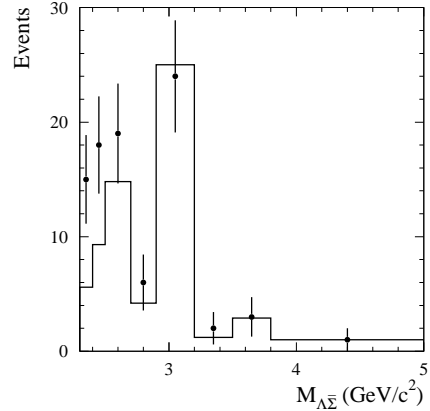}
\caption{The distribution of selected data events
(points with error bars) over chosen mass intervals.
The histogram shows fitted  background.
\label{lamsig_f6}}
\end{figure}
The fit results for $\Lambda\Sigbar^0$ masses below 2.9 GeV/$c^2$ are
shown in Fig.~\ref{lamsig_f5} and summarized in Table~\ref{nlamsig}, 
together with the predictions from the JETSET simulation.
\begin{table*}
\caption{Comparison of the fit results for $\Lambda\Sigbar^0$ masses below 
2.9 GeV/$c^2$ and the predictions from JETSET simulation;
$N_{1s}$,  $N_{0}$, $N_{1b}$, $N_{2}$ and 
$N_{\Lambda\Sigbar^0\pi^0}$ are the
fitted numbers of signal, zero-, one-, two-$\Sigma^0$, and $\Sigma^0\Sigbar^0\pi^0$ 
background events with $\chi^2_{\Lambda\Sigma} < 20$, respectively.
\label{nlamsig}\\}
\begin{ruledtabular}
\begin{tabular}{lccccc}
      &$N_{1s}$     &$N_{2}$      &$N_{0}$      & $N_{1b}$    &$N_{\Lambda\Sigbar^0\pi^0}$\\
data  &$24.1\pm8.4$ &$13.8\pm4.4$ &$17.0\pm7.8$ & $< 5 $      &$3.1\pm2.2$ \\   
JETSET&$50\pm6$     &$17\pm3$     &$3.0\pm1.5$  & $0.6\pm0.6$ &$1.2\pm0.9$ \\
\end{tabular}
\end{ruledtabular}
\end{table*}

The fitting procedure was performed in eight $\Lambda\Sigbar^0$ mass intervals, and
the resulting data distribution is compared to the fitted background in 
Fig.~\ref{lamsig_f6}. An excess of signal events
over background is seen only for $\Lambda\Sigbar^0$ masses below 2.9 GeV/$c^2$.
The number of signal events in each mass interval is listed in Table~\ref{lamsigt};
90\% CL upper limits are given for the intervals with $M_{\Lambda\Sigbar} > 2.9$ GeV/$c^2$.
The significance of the observation of $\Lambda\Sigbar^0$ production in the mass 
region below 2.9 GeV/$c^2$ is 3.3$\sigma$. 

\subsection{Cross section and form factor}\label{crosssec2}
     The cross section for $e^+e^-\to \Lambda\Sigbar^0$ is calculated from
the $\Lambda\Sigbar^0$ mass spectrum according to Eqs.(\ref{ISRcs}-\ref{ISRlum}).

The detection efficiency is determined from MC simulation and then
corrected for data-MC simulation differences in detector response.
The model dependence of the detection efficiency due to the 
unknown $|G_E/G_M|$ ratio is estimated to be 5\%.
The efficiency corrections summarized in Table~\ref{tab_ef2_cor} were
discussed in Secs.~\ref{efflam} and \ref{crosssec1}.
\begin{table}
\caption{The values of the various efficiency corrections for 
the process $e^+e^-\to \Lambda\Sigbar^0\gamma$.
\label{tab_ef2_cor}\\}                                
\begin{ruledtabular}
\begin{tabular}{ll}
effect                &$\delta_i$, (\%) \\
\hline
$\chi^2_{\Lambda\Sigma} < 20$        & $-2.0\pm6.0$  \\
track reconstruction                 & $-1.0\pm3.8$  \\
$\antiproton$ nuclear interaction        & $+1.0\pm0.4$  \\
PID                                  & $+0.6\pm0.6$  \\
photon inefficiency                  & $-2.6\pm0.6$  \\
photon conversion                    & $+0.8\pm0.4$  \\
\hline                                                                                                                    
total                                & $-3.2\pm7.2$ \\
\end{tabular}                                                                                                            
\end{ruledtabular}
\end{table}        
\begin{table*}
\caption{The $\Lambda\Sigbar^0$ invariant mass interval ($M_{\Lambda\Sigbar}$),
net number of signal events ($N_s$),
detection efficiency ($\varepsilon$), ISR luminosity ($L$),
measured cross section ($\sigma$), and effective form factor ($F$) 
for $e^+e^-\to \Sigma^0\Sigbar^0$. The quoted errors on $\sigma$
are statistical and systematic. For the form factor, the total error
is listed.
\label{lamsigt}\\}
\begin{ruledtabular}
\begin{tabular}{ccccccc}
$M_{\Lambda\Sigbar}$ & $N_s$ & $\varepsilon$ &   $L$    & $\sigma$ & $|F|$ \\
(GeV/$c^2$)    &       &               & (pb$^{-1}$) &  (pb)    &       \\
\hline\\
2.308--2.400&$ 9.4_{-4.1}^{+4.6}$  &$0.035\pm0.004$&  5.70&$ 47^{+23}_{-21}\pm5$     &$0.102^{+0.023}_{-0.027}$\\
2.400--2.500&$ 8.7_{-4.7}^{+5.1}$  &$0.041\pm0.004$&  6.52&$ 32^{+19}_{-18}\pm4$     &$0.068^{+0.018}_{-0.022}$\\
2.500--2.700&$ 4.2_{-4.4}^{+4.8}$  &$0.042\pm0.004$& 14.02&$7.1^{+8.2}_{-7.5}\pm0.7$ &$0.029^{+0.014}_{-0.029}$  \\
2.700--2.900&$ 1.8_{-2.3}^{+2.7}$  &$0.041\pm0.004$& 15.38&$2.9^{+4.3}_{-3.7}\pm0.3$ &$0.018^{+0.011}_{-0.018}$   \\
2.900--3.300&$ < 9.0$              &$0.040\pm0.004$& 25.78&$   < 8.7$                &$< 0.033$                \\
3.300--3.500&$ < 5.3$              &$0.041\pm0.005$& 29.28&$   < 4.5$                &$< 0.025$                \\
3.500--3.800&$ < 5.1$              &$0.038\pm0.005$& 33.13&$   < 4.1$                &$< 0.026$                \\
3.800--5.000&$ < 3.5$              &$0.034\pm0.004$&180.38&$   < 0.6$                &$< 0.011$                \\
\end{tabular}
\end{ruledtabular}
\end{table*}
The corrected detection efficiencies are listed in Table~\ref{lamsigt}.
The uncertainty in efficiency takes into account simulation statistical
error,  model uncertainty, the error on the $\Lambda \to \proton\pi^-$ branching
fraction, and the uncertainty in the efficiency correction. 

The measured values of the $e^+e^-\to \Lambda\Sigbar^0$
cross section are listed in Table~\ref{lamsigt}, together with
those of the effective form factor.
\footnote{For the $e^+e^-\to \Lambda\Sigbar^0$ process, 
Eq.(\ref{eq4}) must be modified by the substitutions 
$\beta=(1-(m_\Lambda-m_\Sigma)^2/m^2)\,2P_\Lambda^\ast/m$
and $\tau=m^2/(m_\Lambda+m_\Sigma)^2$~\cite{mil2},
where $P_\Lambda^\ast$ is the baryon momentum.}
The quoted cross section errors are statistical and
systematic. The latter includes systematic uncertainty in
detection efficiency, the error on the total integrated luminosity
(1\%), and the radiative correction uncertainty (1\%).
This is the first measurement of the $e^+e^-\to \Lambda\Sigbar^0$
cross section. The upper limit set by DM2~\cite{DM2ll} at
2.386 GeV ( $< 75$ pb) is consistent with our results.

Assuming that all events in the 2.90--3.30 GeV/$c^2$ mass range
result from $J/\psi\to \Lambda\Sigbar^0$ decay we obtain an
upper limit for the $J/\psi\to \Lambda\Sigbar^0$ branching fraction
${\cal B}(J/\psi\to \Lambda\Sigbar^0) < 2\times 10^{-4}$,
which is slightly higher than the only other estimate,
${\cal B}(J/\psi\to \Lambda\Sigbar^0) < 1.5\times 10^{-4}$~\cite{mark1}.

\section{Summary}
\begin{figure}
\includegraphics[width=.4\textwidth]{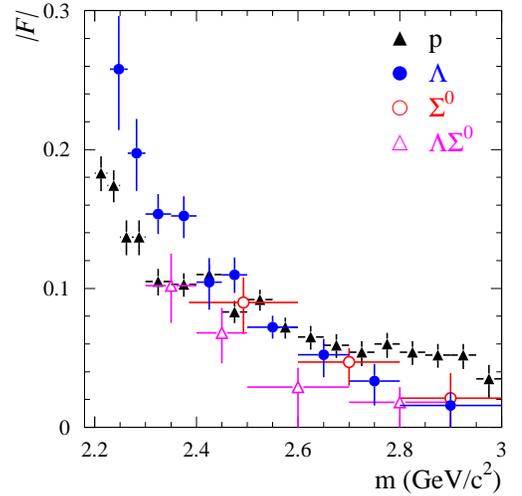}
\caption{
The measured dependence of the baryon form factors 
on dibaryon invariant mass. The proton data are taken from
Ref.~\cite{BADpp}.
\label{ffall}}
\end{figure}
The processes $e^+e^-\to \Lambda\Lbar\gamma$, 
$\Lambda\Sigbar^0\gamma$, and $\Sigma^0\Sigbar^0\gamma$
have been studied for dibaryon invariant mass up to 5 GeV/$c^2$.
From the measured dibaryon mass spectra we obtained the
$e^+e^-\to \Lambda\Lbar$, $\Lambda\Sigbar^0$, and $\Sigma^0\Sigbar^0$
cross sections and baryon effective form factors. 
Our results on the measurements of the various baryon form factors
for dibaryon invariant masses above $\Lambda\Lbar$ threshold
are shown in Fig.~\ref{ffall}. 

For $e^+e^-\to \Lambda\Lbar\gamma$  
we analyzed the $\Lambda$ angular distributions 
in the mass range from threshold to 2.8 GeV/$c^2$
and extracted the $|G_E/G_M|$ ratio. Our results are
\begin{eqnarray}
|G_E/G_M|=1.73^{+0.99}_{-0.57}\;\;\mbox{ for 2.23--2.40 GeV}/c^2,\nonumber\\
|G_E/G_M|=0.71^{+0.66}_{-0.71}\;\;\mbox{ for 2.40--2.80 GeV}/c^2,\nonumber
\end{eqnarray}
and are consistent both with
$|G_E/G_M|=1$ and with the results
for $e^+e^-\to \proton\antiproton$~\cite{BADpp},
where this ratio was found to be significantly
greater than unity near threshold.

The measurement of the $\Lambda$ polarization 
enables the extraction of
the relative phase between the $\Lambda$ electric 
and magnetic form factors.
The limited statistics of the present experiment 
allow us to set only
very weak limits on this phase:
$$-0.76<\sin{\phi}<0.98.$$

From the events in the $J/\psi$ and $\psi(2S)$ regions
the products,
\begin{eqnarray}
\Gamma(J/\psi\to e^+e^-){\cal B}(J/\psi\to \Lambda\Lbar) = 
(10.7\pm 0.9\pm 0.7)\mbox{ eV},\nonumber\\
\Gamma(J/\psi\to e^+e^-){\cal B}(J/\psi\to \Sigma^0\Sigbar^0)=
(6.4\pm 1.2\pm 0.6)\mbox{ eV},\nonumber\\
\Gamma(\psi(2S)\to e^+e^-){\cal B}(\psi(2S)\to \Lambda\Lbar) = 
(1.5\pm0.4\pm0.1)\mbox{ eV},\nonumber
\end{eqnarray}
have been measured, and, using the known $e^+e^-$ partial
widths,
the corresponding branching ratios have been obtained:
\begin{eqnarray}
{\cal B}(J/\psi\to \Lambda\Lbar)=(1.92\pm0.21)\times 10^{-3},\nonumber\\
{\cal B}(J/\psi\to \Sigma^0\Sigbar^0)=(1.16\pm0.26)\times 10^{-3},\nonumber\\
{\cal B}(\psi(2S)\to \Lambda\Lbar)=(6.0\pm1.5)\times 10^{-4}.\nonumber
\end{eqnarray}

\begin{acknowledgments}
We thank A.~I.~Milstein
for useful discussions.
We are grateful for the                                                 
extraordinary contributions of our \pep2\ colleagues in                 
achieving the excellent luminosity and machine conditions               
that have made this work possible.                                      
The success of this project also relies critically on the               
expertise and dedication of the computing organizations that            
support \babar.                                                         
The collaborating institutions wish to thank                            
SLAC for its support and the kind hospitality extended to them.         
This work is supported by the                                           
US Department of Energy                                                 
and National Science Foundation, the                                    
Natural Sciences and Engineering Research Council (Canada),             
the Commissariat \a l'Energie Atomique and                             
Institut National de Physique Nucl\'eaire et de Physique des Particules 
(France), the                                                           
Bundesministerium f\"ur Bildung und Forschung and                       
Deutsche Forschungsgemeinschaft                                         
(Germany), the                                                          
Istituto Nazionale di Fisica Nucleare (Italy),                          
the Foundation for Fundamental Research on Matter (The Netherlands),    
the Research Council of Norway, the                                     
Ministry of Science and Technology of the Russian Federation,           
Ministerio de Educaci\'on y Ciencia (Spain), and the                    
Science and Technology Facilities Council (United Kingdom).             
Individuals have received support from                                  
the Marie-Curie IEF program (European Union) and                        
the A. P. Sloan Foundation.                   
\end{acknowledgments} 
\newpage
\appendix*
\begin{widetext}
\section{\boldmath Angular distributions and $\Lambda$ polarization
in the $e^+e^-\to \Lambda\Lbar\gamma$ 
reaction}

The formulae given in this section are taken from Ref.~\cite{dkm}.
The process $e^+e^-\to \Lambda\Lbar\gamma$ is considered
in the $e^+e^-$ center-of-mass frame, where the electron
has momentum {\boldmath $p$} and energy $\varepsilon$,
and the photon has momentum {\boldmath $k$} and energy $\omega$.
The $\Lambda$ momentum {\boldmath $P$} is given in the
$\Lambda\Lbar$ rest frame. The differential cross section
summed over the polarization of one of the final particles
is given by
$${\rm d}\sigma=\frac{\alpha^3 P {\rm d}^3k {\rm d}\Omega_\Lambda}
{16\pi^2\omega\varepsilon^2Q^3[1-({\mathbf n}\cdot \mbox{\boldmath$\nu$})^2]}{\cal A}
(1+\mbox{\boldmath$\zeta_f$}\cdot{\mathbf s}),\;\;
{\cal A}=2|G_M|^2(1+N^2)+                     
\left( \frac{4m_\Lambda^2}{Q^2}|G_E|^2-|G_M|^2 \right)
([{\mathbf n}\times{\mathbf f}]^2+[{\mathbf N}\times{\mathbf f}]^2),$$
$$\mbox{\boldmath$\zeta_f$}=\frac{4m_\Lambda}{Q{\cal A}}\mbox{Im}(G_E^\ast G_M)
\left(                                                      
({\mathbf n}\cdot{\mathbf f})[{\mathbf n}\times{\mathbf f}]+
({\mathbf N}\cdot{\mathbf f})[{\mathbf N}\times{\mathbf f}] 
\right);\;\;\;\;               
{\mathbf n}=\frac{{\mathbf k}}{\omega},\;                 
\mbox{\boldmath$\nu$}=\frac{{\mathbf p}}{\varepsilon},\;    
{\mathbf N}=\frac{\mbox{\boldmath$\nu$}+                    
(\gamma-1)({\mathbf n}\cdot \mbox{\boldmath$\nu$}){\mathbf n}}
{\sqrt{\gamma^2-1}},$$
$$N^2=({\mathbf n}\cdot \mbox{\boldmath$\nu$})^2+
\frac{1}{\gamma^2-1},\;                                       
\gamma=\frac{2\varepsilon-\omega}{Q},\;
Q=\sqrt{\varepsilon(\varepsilon-\omega)},\;
P=|{\mathbf P}|=\sqrt{Q^2/4-m_\Lambda^2},\;
{\mathbf f}=\frac{{\mathbf P}}{P}.         
$$
Here ${\mathbf s}$ and {\boldmath$\zeta_f$} are the spin and polarization
vectors of the $\Lambda$ in its rest frame. 
\end{widetext}

\end{document}